\documentclass[numberedappendix]{emulateapj}

\usepackage{natbib}
\usepackage{amsmath}
\usepackage{graphicx}

\usepackage{color}
\usepackage[normalem]{ulem}
\usepackage{url}

\DeclareUrlCommand\ULurl{%
  \renewcommand\UrlLeft{\bgroup}%
  \renewcommand\UrlRight{\egroup}}

\newcommand{\mintext}{\text{min}}
\newcommand{\maxtext}{\text{max}}
\newcommand{\rad}{\text{rad}}
\newcommand{\IR}{\text{IR}}
\newcommand{\ir}{\text{IR}}

\newcommand{\shot}{\text{shot}}

\newcommand{\fov}{\text{FOV}}

\newcommand{\Fb}{\mathbf{F}}
\newcommand{\Mb}{\mathbf{M}}
\newcommand{\Cb}{\mathbf{C}}
\newcommand{\Ab}{\mathbf{A}}
\newcommand{\xb}{\mathbf{x}}
\newcommand{\pb}{\mathbf{p}}
\newcommand{\qb}{\mathbf{q}}
\newcommand{\Wb}{\mathbf{W}}         

\shorttitle{Foreground and sensitivity analysis for broad band (2D) 21\,\MakeLowercase{cm}--L\MakeLowercase{y}$\alpha$ and 21\,\MakeLowercase{cm}--H$\alpha$ correlation experiments}
\shortauthors{Neben et al.}

\begin{document}

\title{Foreground and sensitivity analysis for broad band (2D) 21\,\MakeLowercase{cm}--L\MakeLowercase{y}$\alpha$ and 21\,\MakeLowercase{cm}--H$\alpha$ correlation experiments probing the Epoch of Reionization}


\author{Abraham R. Neben\altaffilmark{1,2},
Brian Stalder\altaffilmark{3},
Jacqueline N. Hewitt\altaffilmark{1,2},
John L. Tonry\altaffilmark{3}}

\affil{\altaffilmark{1}MIT Kavli Institute, Massachusetts Institute of Technology, Cambridge, MA 02139}
\affil{\altaffilmark{2}Department of Physics, Massachusetts Institute of Technology, Cambridge, MA 02139}
\affil{\altaffilmark{3}Institute for Astronomy, University of Hawaii, 2680 Woodlawn Drive, Honolulu, HI 96822}


\begin{abstract}

A detection of the predicted anticorrelation between $21$\,cm and either Ly$\alpha$ or H$\alpha$ from the Epoch of Reionization (EOR) would be a powerful probe of the first galaxies. While 3D intensity maps isolate foregrounds in low $k_\parallel$ modes, infrared surveys cannot yet match the field of view and redshift resolution of radio intensity mapping experiments. In contrast, 2D (i.e., broad band) infrared intensity maps can be measured with current experiments and are limited by foregrounds instead of photon or thermal noise. We show 2D experiments can measure most of the 3D fluctuation power at $k<0.2$\,Mpc$^{-1}$ while preserving its correlation properties. However, we show foregrounds pose two challenges: (1) simple geometric effects produce percent-level correlations between radio and infrared fluxes, even if their luminosities are uncorrelated; and (2) radio and infrared foreground residuals contribute sample variance noise to the cross spectrum. The first challenge demands better foreground masking and subtraction, while the second demands large fields of view to average away uncorrelated radio and infrared power. Using radio observations from the Murchison Widefield Array and near-infrared observations from the Asteroid Terrestrial-impact Last Alert System, we set an upper limit on residual foregrounds of the 21\,cm--Ly$\alpha$ cross power spectrum at $z\sim7$ of $\Delta^2<181$\,$(\text{kJy/sr}\cdot \text{mK})$ (95\%) at $\ell\sim800$. We predict levels of foreground correlation and sample variance noise in future experiments, showing that higher resolution surveys such as LOFAR, SKA-LOW, and the Dark Energy Survey can start to probe models of the 21\,cm--Ly$\alpha$ EOR cross spectrum. 
\end{abstract}


\section{Introduction}

Deep radio and infrared observations are nearing detection of the first stars and galaxies from the cosmic dawn. As such sources form, they are thought to blow out ionized bubbles, eventually merging and reionizing the universe. See \citet{FurlanettoReview,miguelreview,PritchardLoebReview,Mesinger16} for reviews. First generation 21\,cm observatories such the Murchison Widefield Array (MWA) \citep{tingay13,mwascience} and the Precision Array for Probing the Epoch of Reionization (PAPER) \citep{parsons14,ali15,PoberPAPER64Heating,DannyMultiRedshift} are setting ever tighter limits on redshifted neutral hydrogen emission from the neutral regions between these bubbles, and the now-underway Hydrogen Epoch of Reionization Array (HERA) \citep{deboer16,neben16b,ewallwice16,patra16} is expected to detect and characterize the EOR power spectrum in the coming years. Similar efforts are underway by the Low Frequency Array \citep{lofareorpaper,lofar}. Ultimately, the Square Kilometer Array (SKA)  \citep{ska,ska1,ska2,ska3} will image the EOR over redshift, revealing the detailed hydrogen reionization history of the universe. 

At the same time, new galaxy surveys are beginning to constrain the reionizing sources themselves. Deep galaxy surveys \citep{Bowler2017,Roberts-Borsani2016,Bowler2015,Wilkins12015,Bouwens2015,Bouwens2013,Dunlop2013,Illingworth2013,Ellis2013,Robertson2013,Oesch2013,Grogin2011,Bouwens2011} and cluster lensing surveys \citep{Livermore2016,Ebeling2016,McLeod2016,Bouwens2016,Bouwens2016a,Coe2015,Huang2015,McLeod2015,Atek2015} are finding hundreds of galaxy candidates at $6<z<10$ down to UV magnitudes of $M_{AB}\sim-17$ \citep{Finkelstein2015}, and extremely wide surveys are searching for the rare bright galaxies \citep{Bernard2016,Calvi,Schmidt2014,Bradley2012,Trenti2011} from the reionization epoch. However, current models require the ionizing contribution of far fainter galaxies down to $M_{AB}\sim-13$ \citep{Bouwens2016b,Alvarez2012} in order for reionization to be complete by the time that CMB optical depth measurements \citep{planck16} say it must be. Deeper observations with the James Webb Space Telescope (JWST) \citep{Gardner2006}, expected to probe down to $M_{AB}\sim-15.5$ in order hundred-hour integrations \citep{Finkelstein2015}, will be needed to begin to probe this crucial faint population directly.

Infrared intensity mapping offers several advantages compared to galaxy surveys. Power spectrum analyses can be sensitive to an EOR component even if the signal-to-noise in individual pixels is small, and instead of being limited to the brightest galaxies, intensity mapping is sensitive to the cumulative light from \textit{all} sources. The expected bright Ly$\alpha$ \citep[e.g.][]{primevaltwins} and H$\alpha$ \citep[e.g.][]{brightemissionlines} radiation from EOR galaxies at $z\sim6-8$ is motivating intensity mapping at micron-scale wavelengths. Working around foregrounds is challenging, though. While early studies suggested angular fluctuations in infrared intensity maps traced EOR galaxies \citep[e.g.,][]{kash1,kash2,kash3}, \citet{kash4} find that given current constraints on the EOR, this is unlikely. Intrahalo light \citep{cooray12,zemcov14} and Galactic dust \citep{yue16} have been proposed as more likely explanations for the larger than expected fluctuations, though \citet{mw15} show that much of this excess can be removed with higher resolution measurements using Hubble. In any case, all these components are likely present at some level and degeneracies coupled with imperfect foreground knowledge make isolating the EOR contribution difficult.

For these reasons, cross correlation with 21\,cm maps may be the cleanest way to extract the EOR component of the near infrared background. The synergy is clear: the galaxies sourcing reionization generate strong Ly$\alpha$ emission, while the neutral regions between them glow at rest frame 21\,cm. On typical ionized bubble scales, bright spots in IR maps likely correspond to ionized regions, and thus, 21\,cm dark spots, and vice versa. This effect is expected to source an anticorrelation, seen in simulations by \citet{silva12,Heneka2016} and modeled analytically by \citet{feng17,mao14}.

A similar large scale anticorrelation is found by \citet{lidz09,park14} in simulations of 21\,cm cross correlation with galaxy redshift surveys. However, conducting redshift surveys both wide and deep enough to cross correlate with 21\,cm maps is challenging due to the hugely different spatial scales probed by 21\,cm experiments and spectroscopic galaxy surveys. For instance, the $\sim3'$ angular resolution of the MWA is nearly equal to the field of view of the Hubble Deep Field and the James Webb Space Telescope (JWST). In a source-by-source manner, \citep{beardsley15} show that this anticorrelation may be studied by inspecting the 21\,cm brightness temperatures at the locations of JWST sources, but as discussed above, source detections will be limited to the brightest and rarest objects.

Intensity mapping thus holds great promise to extend EOR science, and 3D intensity mapping (i.e., with a redshift dimension) has the advantage of avoiding the majority of continuum emission from intermediate redshift galaxies. This foreground emission is expected to contaminate only the lowest few line of sight Fourier modes \citep{gong17}, while line interlopers at intermediate redshifts are easily masked \citep{Gong2014,gong17,pullen14,Comaschi16}. 3D power spectra are also easier to understand theoretically as they quantify emission from a fundamentally 3D volume, and early demonstrations of this type of analysis are given by \citet{Chang2010,Masui2013} who detected the cross correlation between 21\,cm emission and a galaxy redshift survey at $z\sim1$. However, near infrared intensity mapping in 3D likely requires space-based observations to avoid atmospheric OH lines \citep[e.g.][]{sullivan12}, as well as fine spectral resolution to match the redshift resolution of typical 21\,cm experiments. 

For instance, \citet{PoberNextGen} show that with moderate foreground assumptions, HERA should achieve $>5\sigma$ detections of the 21\,cm power spectrum 	over $0.2<k_\parallel<0.5$ at $z\sim8$, corresponding to redshift scales of $0.03<\Delta z<0.1$. Resolving these same line of sight modes of the Ly$\alpha$ field at the same redshift requires a spectral resolution\footnote{The redshift resolution of a spectral line intensity mapping experiment observing emission at redshift $z$ is given by $\Delta z=(1+z)/R$, where $R$ is the spectral resolving power.} of $80<R<250$. Achieving this high spectral resolution over a large enough field of view to match a 21\,cm survey is challenging. The proposed SPHEREx mission \citep{ScienceWithSpherex,SpherexWhitePaper} would image the entire sky in the near infrared with $R=40$ spectroscopy for a cost of roughly \$100M, and the concept Cosmic Dawn Intensity Mapper \citep{cooray16} mission would achieve R=200--300 over 10 square degrees for nearly ten times the cost.

In contrast, 2D (i.e, broad band) intensity mapping enables similar science with far shallower and cheaper observations \citep{StarsAndReionization,mao14}, though here the main challenge is imperfect foreground removal. Even if radio and infrared foreground residuals are uncorrelated, they leak power into the cross correlation analysis which averages down only over sufficiently large fields of view. As OH emission from the atmosphere is relatively smooth over few degree scales \citep{high10}, there is hope that these observations can be conducted from the ground for a further reduction in cost. A number of new ground-based wide field surveys are coming online such as the Dark Energy Survey \citep{des16}, Pan-STARRS \citep{tonry12}, and the Asteroid Terrestrial-impact Last Alert System (ATLAS) \citep{tonry11}. Further, the Transiting Exoplanet Survey Satellite \citep{ricker14} will survey the entire sky over 600--1000\,nm band, and the Wide-Field Infrared Survey Telescope (WFIRST) \citep{Spergel2013} will observe instantaneous deep fields 100$\times$ larger those of Hubble and JWST. It is important to note that a large, uniform focal plane greatly facilitates intensity mapping, lest structures on relevant angular scales be lost in the calibration of many independent regions of a segmented focal plane, such as that of Pan-STARRS. 

In this paper we study the foregrounds in broad band 21\,cm--Ly$\alpha$ and 21\,cm--H$\alpha$ intensity mapping correlation experiments targeting the EOR. We begin in Sec. 2 with a review of our Fourier transform and power spectrum conventions.
  In Sec. 3 we present the MWA and ATLAS observations we use and discuss processing these data into images. 
   In Sec. 4 we characterize the bright radio and infrared point source foregrounds and 
   demonstrate that geometric effects introduce slight positive correlations which will overpower the cosmological signal
   unless significant masking and subtraction are conducted. 
   In Sec. 5, we study how best to mask and subtract radio and infrared foregrounds in real world images and 
   quantify the foreground residuals. We set the first limits 
   on residual foregrounds of the the broad band 21\,cm--Ly$\alpha$ cross spectrum at $z\sim7$ using data from the MWA
     and ATLAS, and predict the sensitivities of future experiments and compare them with the expected levels of geometric foreground correlation, 
     illustrating what it will take to realize this measurement.

\section{Power spectrum and correlation conventions}
\label{sec:pspecconventions}

\subsection{Power spectrum definitions}
\label{sec:pspecdefs}

We define the 3D power spectrum $P(\vec{k})$ of the image cube $I(\vec{x})$ as 
\begin{equation}
\label{eqn:pspec3Ddef}
	P(\vec{k}) = \frac{\langle|\tilde{I}(\vec{k})|^2\rangle}{V}
\end{equation}
where $\tilde{I}(\vec{k})$ is given by
\begin{equation}
	\tilde{I}(\vec{k})=dV\sum_{\vec{x}}I(\vec{x})e^{i\vec{k}\cdot\vec{x}},
\end{equation}
Here $V$ is the survey volume, and $dV$ is the voxel size. Note that $P(\vec{k})$ has units of $[I]^2\cdot\text{Mpc}^3$, and that we often plot instead the more intuitive quantity $\Delta(k)=[k^3 P(k)/2\pi^2]^{1/2}$, where $P(k)$ is the average of $P(\vec{k})$ within the 1D power spectrum bin $k$.

Similarly, over narrow fields of view, the angular power spectrum $C(\vec{\ell})$ of a 2D (e.g, broad band) image $I(\vec{\theta})$ can be shown to be approximately
\begin{equation}
\label{eqn:Cldef0}
	C(\vec{\ell}) = \frac{\langle|\tilde{I}(\vec{\ell})|^2\rangle}{\Omega} 
\end{equation}
where $\tilde{I}(\vec{\ell})$ is given by
\begin{equation}
	\tilde{I}(\vec{\ell})=d\Omega\sum_{\vec{\theta}}I(\vec{\theta})e^{i\vec{\ell}\cdot\vec{\theta}},
\end{equation}
where $\Omega$ is the survey solid angle, and $d\Omega$ is the pixel size. Thus, over a narrow field of view, we need only evaluate a Fourier transform to estimate the angular power spectrum. Writing this out in detail gives\footnote{Note that the normalization of $d\theta^2/N^2$ has been misstated as $1/N^2$ by \citet{zemcov14} (Eqn. 3 of their supplement) and $d\theta^2$ by \citet{cooray12} (Eqn. 1 of their supplement).} 
\begin{equation}
\label{eqn:Cldef}
	C(\vec\ell(a,b))=\left|\sum_{m,n}I(m,n)\exp\left(-\frac{2\pi i}{N}  (am+bn)\right)\right|^2\frac{d\theta^2}{N^2}
\end{equation}
where $d\theta=d\theta_x=d\theta_y$ is the pixel size, $N\equiv N_x=N_y$ is the number of pixels on a side of a square image, and $-N/2 \leq a,b\leq N/2$ are integers. Further $\ell^2=\ell_x^2+\ell_y^2$, and 
\begin{eqnarray}
\ell_x&=&2\pi a/N d\theta \label{eqn:elldef}\\
\ell_y&=&2\pi b/Nd\theta \label{eqn:elldef2}
\end{eqnarray}
Note that $C(\vec{\ell})$ has the units of $[I]^2\cdot d\theta^2$, and we often work with $\Delta(\ell)=[\ell(\ell+1)C(\ell)/2\pi]^{1/2}$ which has the same units\footnote{Over small fields of view, we must necessarily work at large $\ell$, implying that $\ell(\ell+1)\approx\ell^2$, which has units of inverse steradians in view of Eqns. \ref{eqn:elldef} and \ref{eqn:elldef2}} as I. Here $C(\ell)$ is the average of $C(\vec\ell)$ within the 1D power spectrum bin $\ell$.

The 3D 21\,cm power spectrum is often cylindrically binned from 3D $\vec{k}$ space to 2D $(k_\perp,k_\parallel)$ space where $k_\perp^2\equiv k_x^2+k_y^2$ represents modes perpendicular to the line of sight, and $k_\parallel=k_z$ represents modes along the line of sight. We show in Appendix \ref{sec:pspecrelation} that this cylindrically binned power spectrum is related to the angular power spectrum of a broad band image (over a narrow field of view) as
\begin{equation}
\label{eqn:convertP3toP2}
P(k_\perp,k_\parallel=0)=D_c^2 \Delta D_c C(\ell(k_\perp)).
\end{equation}
Here $\ell=D_c k_\perp$, where $D_c$ is the comoving line of sight distance to the center of the cube, and $\Delta D_c$ is the comoving depth of the cube.

\subsection{Cross spectrum vs. coherence}

The 3D and 2D cross spectra are defined, extending Eqns. \ref{eqn:pspec3Ddef} and \ref{eqn:Cldef0} to the cross spectrum as
\begin{eqnarray}
	P_{12}(\vec{k}) &=& \frac{\langle\tilde{I}_1^*(\vec{k})\tilde{I}_2(\vec{k})\rangle}{V}\\
	C_{12}(\vec{\ell}) &=& \frac{\langle \tilde{I}_1^*(\vec{\ell})\tilde{I}_2(\vec{\ell})\rangle}{\Omega}
\end{eqnarray}
where $1$ and $2$ denote the 21\,cm and the IR fields, respectively. The cross spectrum is a quantity which ranges between $\pm(C_{1}(\vec{\ell})C_{2}(\vec{\ell}))^{1/2}$ in the 2D case, depending on how correlated, uncorrelated, or anticorrelated the two fields are. It is thus often renormalized as  
\begin{equation}
\label{eqn:Cldefcross}
	c_{12}(\vec{\ell}) \equiv \frac{C_{12}(\vec{\ell}) }{\sqrt{C_1(\vec{\ell})  C_2(\vec{\ell}) }}
\end{equation}
where $c$ is known as the coherence and is insensitive to a simple rescaling of either field. However, uncorrelated foreground residuals in either field will substantially bias the coherence towards zero \citep{lidz09,furlanettolidz07}, whereas they merely contribute a zero mean noise to the cross spectrum. Slight foreground correlations, of course, will bias the cross spectrum as well, as we explore later. 

\section{Observations and Imaging}
\subsection{21\,cm Observations}
\label{sec:mwaobservations}

The MWA is a low frequency radio interferometer in Western Australia consisting of 128 phased array tiles, each with $\sim30^\circ\times(150\text{\,MHz}/f)$  beams (full-width-at-half-maximum) and steerable in few degree increments with a delay line beamformer. We use low frequency observations of a quiet field centered at (RA,Dec)=($0^\circ$,$-27^\circ$) J2000 recorded over 30.72\,MHz bandwidth centered at 186\,MHz, corresponding to $z=6.0-7.3$ for rest frame 21\,cm. 

We use MWA image products produced by \citet{beardsley16}. The MWA observations are recorded as 2\,min ``snapshots'' which are flagged for RFI using COTTER \citep{AndreMWARFI}, then calibrated and imaged using Fast Holographic Deconvolution\footnote{\ULurl{https://github.com/EoRImaging/FHD}}. Model visibilities are simulated from a foreground model of diffuse  and point source \citep{PattiCatalog1} emission in the field, and used for both calibration and foreground subtraction. For each snapshot, FHD produces naturally weighted data and model image cubes as well as primary and synthesized beam cubes. FHD outputs these ``cubes'' in HEALPix format per frequency. Note that this processing is performed in parallel on ``odd'' and ``even'' data cubes whose data are interleaved at a 2\,sec cadence for the purpose of avoiding a thermal noise bias in autospectrum analyses. In cross spectrum analyses, the noise between the radio and IR images is independent, so in principle we should average the odd and even cubes together to achieve the lowest noise. However, for simplicity, we use only the even cube in this work given that thermal noise is much smaller than residual foregrounds.

Following \citet{dillonneben}, we rotate these HEALPix maps so the MWA field center lies at the north pole, then project the pixels onto the $xy$ plane to obtain naturally weighted image space cubes of the raw data ($I_\text{nat}$), the model data ($I_\text{nat,mod}$), the synthesized beam ($I_w$) (i.e., the Fourier transform of the $uv$ weights), and the primary beam, all in orthographic projection. We flag the upper and lower 80\,kHz channels in each of 24 coarse channels across the band to mitigate aliasing, average each cube over frequency to make broad band images, then apply uniform weighting using
\begin{equation}
\label{eqn:uniformweighting}
I_\text{uni}(\vec{\theta}) = \frac{10^{-26}\lambda^2}{k_B } \sum_{\vec{u}} \frac{\tilde{I}_\text{nat}(\vec{u})}{\tilde{I}_w(\vec{u})} e^{-2\pi i \vec{\theta}\cdot\vec{u}}d^2u
\end{equation}
where $\tilde{I}_i(\vec{u}) = \sum_{\vec{\theta}} I_i(\vec{\theta}) e^{2\pi i\vec{\theta}\cdot\vec{u}} d\Omega$ for $i=\text{nat},w$. The units of $I_\text{nat}$ and $I_w$ are Jy/sr and 1/sr, respectively, as seen from their approximate definitions of $I_\text{nat}(\vec{\theta})\approx\sum_j V(\vec{u}_j)e^{-2\pi i\vec{u}_j\cdot\vec{\theta}}du^2$ and $I_w(\vec{\theta})\approx\sum_j e^{-2\pi i\vec{u}_j\cdot\vec{\theta}}du^2$, where the sums are over all measured visibilities $V(\vec{u}_j)$ in units of Jy. These definitions are only approximate because FHD performs corrections to account for wide-field effects.

\subsection{IR Observations}

\begin{figure}[h]
\centering
\includegraphics[width=3.3in]{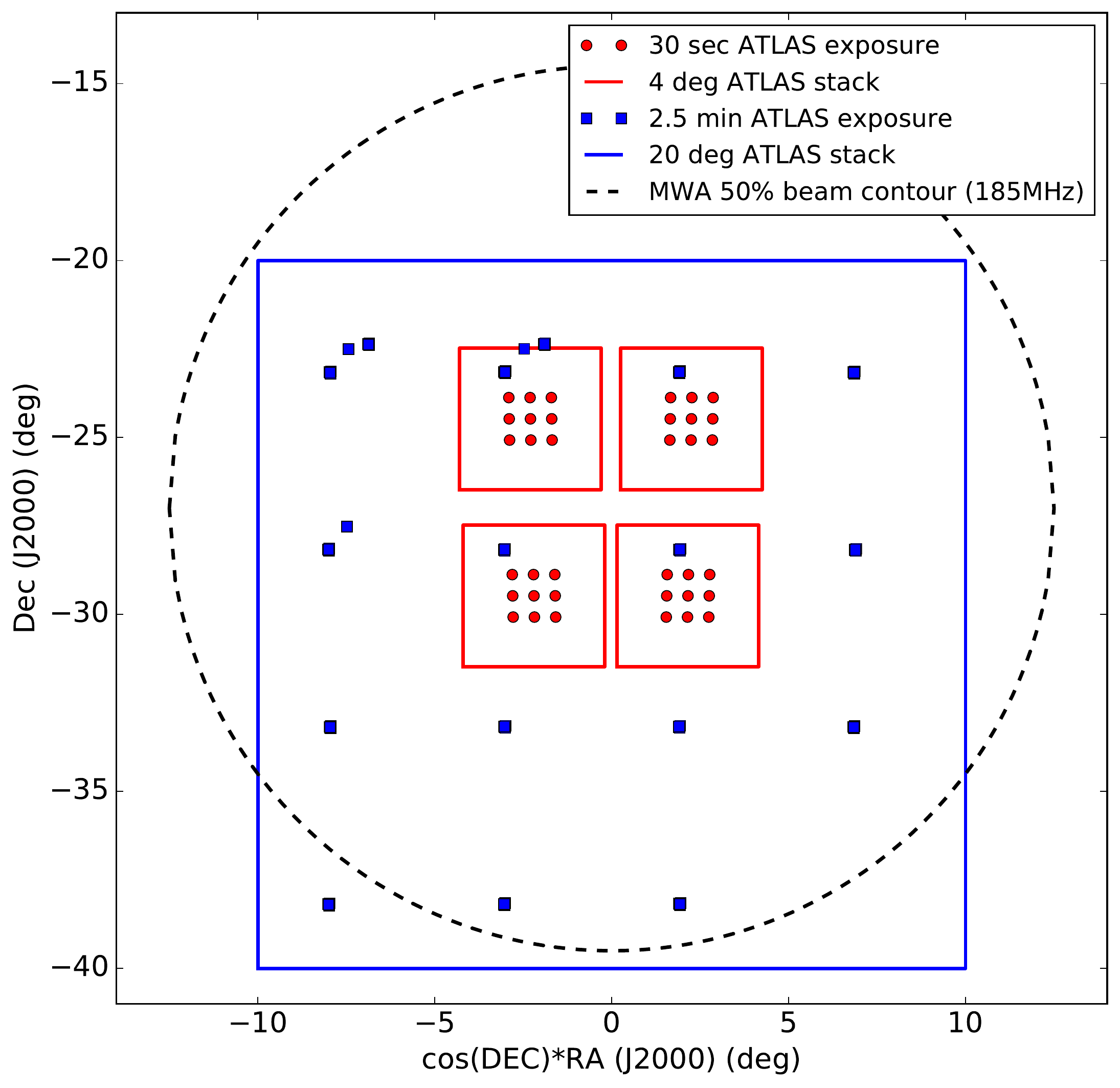}
\caption[Diagram of the MWA deep integration field and our ATLAS surveys.]{The MWA deep integration field (black dashed circle) is shown relative to our two ATLAS surveys. Blue square markers show the observation centers of our wide ATLAS survey aimed at studying foregrounds, and the large blue square outline shows the stacked image. Red circle markers show observation centers for our slightly deeper survey, and red square outlines show the four stacked images we generate. Note that the ATLAS field of view is 5.5$^\circ$.}
\label{fig:surveyoverview}
\end{figure}

ATLAS is a system of multiple 0.5\,m f/2 wide-field telescopes \citep{tonry11} designed for near-earth asteroids detection and tracking. Two telescopes are currently in operation located on the islands of Maui and Hawaii.   Each telescope has a single $10,560\times10,560$ STA1600 CCD sensor, with a pixel scale of $1.86''$, giving a field of view of $5.5^\circ$ on a side.  We observe in the Johnson I band (810\,nm center with 150\,nm full-width-at-half-max) with the Maui telescope, corresponding to $z=5.1-6.3$ for Ly$\alpha$.  While this redshift range doesn't exactly match that of our radio observations, it overlaps sufficiently for our purpose of characterizing the noise and foregrounds in 21cm--Ly$\alpha$ cross correlation experiments. The automatic processing pipeline provides single sky-flattened (using a median of the nightly science exposures) images registered to nominal 2MASS astrometry.

We perform two separate observing campaigns, which we illustrate in Fig. \ref{fig:surveyoverview}. We first perform a wide survey to best characterize bright foregrounds. We raster scan a roughly $20^\circ\times20^\circ$ grid with $5^\circ$ spacing over the MWA field (dashed black circle), integrating for 2.5\,min at each pointing (blue square markers). The observations were conducted between 2016/09/07 22:00  and 2016/09/08 00:50 Hawaii-Aleutian Standard Time, when the moon was 36\% illuminated. We use {\tt swarp}\footnote{\ULurl{http://www.astromatic.net/software/swarp}} \citep{swarp} to stack all these frames over a $20^\circ$ orthographic field centered on (RA,Dec) = $(0^\circ,-30^\circ)$ (blue square) with $1.86''$ resolution, using the default background subtraction settings to mitigate temporal and spatial background variation. 


Our second campaign is a slightly deeper survey designed to better mitigate airglow fluctuations and CCD systematics for the purpose of studying faint foregrounds below the detection limit. We select four 5 deg fields positioned around the MWA beam peak for best cross correlation precision: (RA,Dec) = $(-2.5^\circ, -24.5^\circ)$, $(2.5^\circ, -24.5^\circ)$, $(-2.5^\circ, -29.5^\circ)$, $(2.5^\circ, -29.5^\circ)$ (J2000). We raster scan a $3\times3$ grid of 30\,sec observations within each field (red circle markers) intended to mitigate slight amplifier non-uniformities across the CCD array. The observations were conducted on 2016/11/02 between 21:47 and 23:11 Hawaii-Aleutian Standard Time, when the moon was 5\% illuminated. 

We stack the frames in each of the four deep fields using {\tt swarp} over only the central $4^\circ\times 4^\circ$ region over which all nine 30\,sec frames overlap (red squares). Otherwise slight background discontinuities would be introduced by the different temporal coverage of different regions of the stack. In this stacking we disable background subtraction for the purpose of studying the effects of airglow-induced diffuse backgrounds.

\section{Point source foregrounds}
\label{sec:pointsourcefgs}

In this section we show with data and simulations that geometric effects can introduce slight correlations between radio and infrared foreground fluxes in broadband surveys, and we quantify how these correlations vary with source masking depth. As the foregrounds are so much brighter than the EOR emission, even slight foreground correlations can bury the predicted EOR anti-correlation, though the expected sign difference will help to identify which effect has been detected.

\subsection{Catalogs}
\label{sec:catalogs}

To characterize the bright sources relevant to broad band  21\,cm--Ly$\alpha$ and  21\,cm--H$\alpha$ intensity mapping correlation measurements, we calculate the correlations between catalogs at 185\,MHz, 850\,nm, and 4.5\,$\mu$m as a function of mask depth. These bands correspond roughly to 21\,cm, Ly$\alpha$, and H$\alpha$, respectively, from $z\sim6-7$. 

\begin{figure*}[t]
\centering
\includegraphics[width=6.5in]{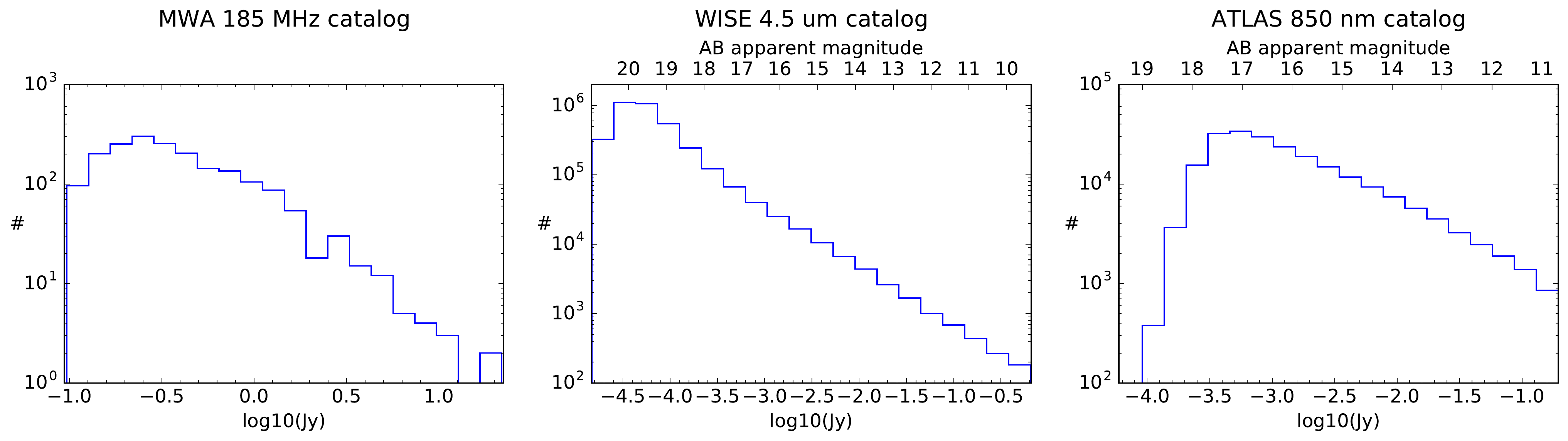}
\caption[Histogram of source fluxes in the 185\,MHz, 850\,nm, and 4.5\,$\mu$m catalogs.]{Histogram of source fluxes in the 185\,MHz catalog (left), the 4.5\,$\mu$m catalog (center), and the 850\,nm catalog (right). The catalogs are complete to roughly 250\,mJy, 0.09\,mJy, and 0.5\,mJy respectively.}
\label{fig:cataloghistograms}
\end{figure*}

We use the 185\,MHz catalog reduced from deep observations of the MWA field depicted in Fig. \ref{fig:surveyoverview} by \citet{PattiCatalog1}. Fig. \ref{fig:cataloghistograms} (left panel) shows a histogram of source fluxes in this field. The survey depth varies somewhat over the MWA field due to the varying primary beam, resulting in a catalog which is 50\% complete down to 150\,mJy and 95\% complete down to 250\,mJy. These completeness levels are shallower than the 70\,mJy completeness quoted by \citet{PattiCatalog1} due to our large rectangular field. The MWA's intrinsic astrometry is at the $2'-3'$ level, though \citep{PattiCatalog1} cross-match with higher frequency catalogs to achieve order $10''$ astrometry. 

We use the W2 band of ALLWISE \citep{Wright2010,allwise} as our 4.5\,$\mu$m catalog. We download the list of sources within the MWA field using the All Sky Search on the NASA/IPAC Infrared Science Archive\footnote{\ULurl{http://irsa.ipac.caltech.edu}}, and plot the histogram of source fluxes in Fig. \ref{fig:cataloghistograms} (center panel). This ALLWISE band is specified to be 95\% complete down to 88\,$\mu$Jy (15.7 AB mag), though it has slight sky coverage non-uniformities due to satellite coverage.

\begin{figure}[t]
\centering
\includegraphics[width=2.75in]{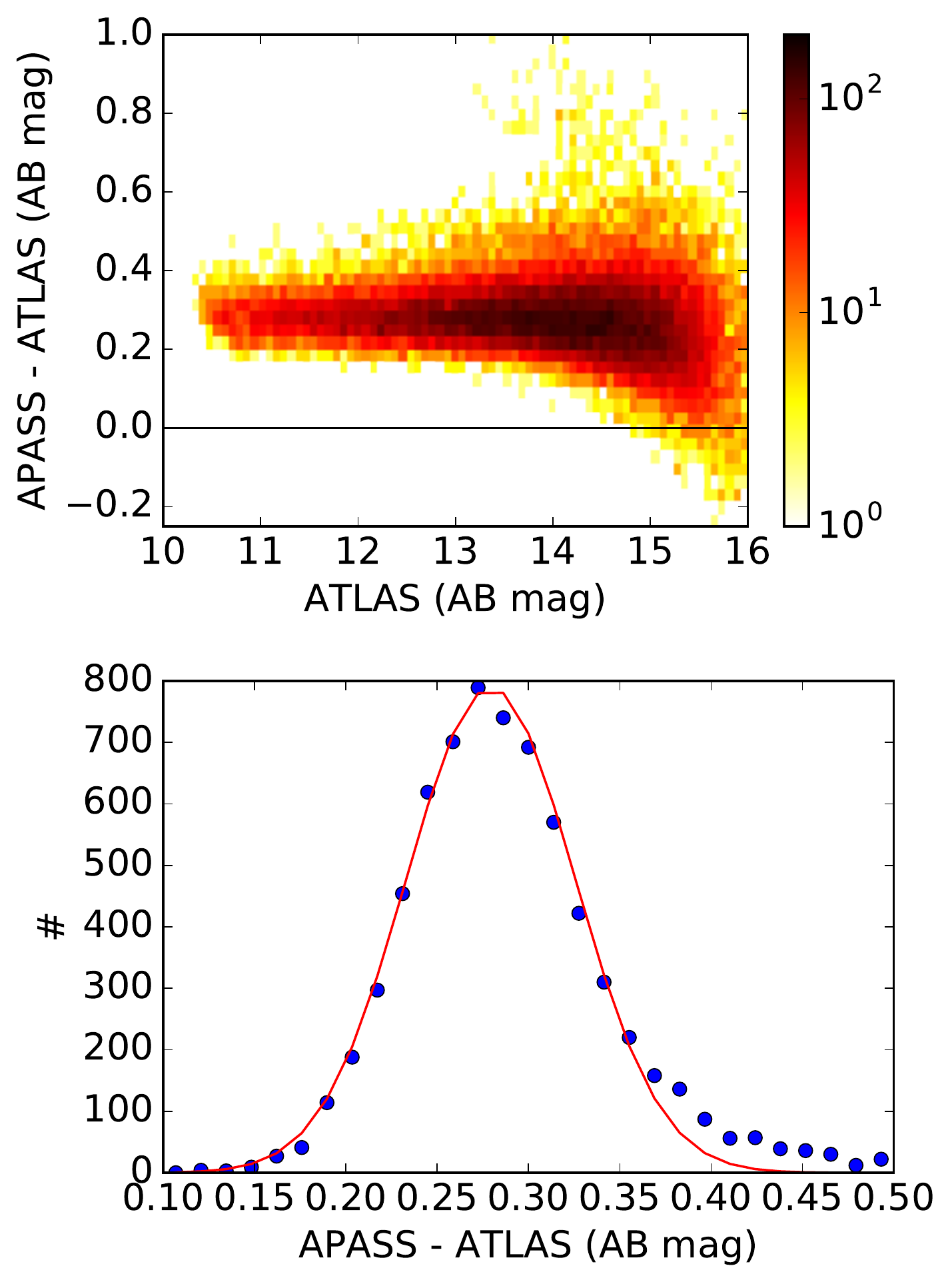}
\caption[To improve the rough initial ATLAS calibration, we cross match ATLAS sources with those from APASS.]{To improve the rough initial ATLAS calibration, we cross match ATLAS sources with those from APASS, and plot a 2D histogram (top) of the relative magnitude as a function of ATLAS magnitude, then fit a Gaussian to the magnitude offset for sources brighter than 13\,mag. We find that our roughly calibrated ATLAS sources are too bright by $0.279\pm0.003$\,mag.}
\label{fig:ATLASvsAPASS}
\end{figure}

Lastly, we run SExtractor\footnote{\ULurl{http://www.astromatic.net/software/sextractor}} \citep{sextractor} on our wide $20^\circ$ ATLAS composite image to generate an 850\,nm catalog. We allow local background bias and noise estimation, set pixel saturation at 20,000 counts to avoid artifacts, and use the {\tt AUTO} aperture profile. We extract sources down to $3\sigma$ above the background, in order to achieve the most complete point source mask. Given that our ATLAS observations have been calibrated and imaged through a preliminary pipeline, we cross match these sources with sources closer than $1''$ in the AAVSO\footnote{American Association of Variable Star Observers} Photometric All Sky Survey\footnote{\ULurl{https://www.aavso.org/download-apass-data}} \citep{apass}, which is complete to $\sim3$\,mJy. We find matches for $\sim20\%$ of ATLAS detections, not unreasonable given the deeper flux limit of ATLAS compared to APASS. Fig. \ref{fig:ATLASvsAPASS} (top) shows a 2D histogram of APASS versus ATLAS magnitude as a function of ATLAS magnitude. We fit a Gaussian to the relative magnitude for sources brighter than 13 mag, and find that our roughly calibrated ATLAS sources are too bright by $0.279\pm0.003$\,mag. Applying this correction, we plot a histogram of ATLAS source fluxes in Fig. \ref{fig:cataloghistograms}) (right panel), finding that our survey is complete to roughly 0.5\,mJy, a factor of 6 deeper than the APASS survey.

\subsection{Catalog radio--infrared flux correlations}
\label{sec:catcorrelations}

Having prepared catalogs of point source foregrounds in our three bands, we proceed to study how they manifest in intensity mapping correlation experiments. Traditionally, radio/infrared correlations have been studied by cross matching high frequency radio detections with infrared sources coincident within a few arcseconds, then plotting radio versus infrared luminosity. Such studies have revealed the well known radio--far infrared correlation thought to be due to massive star formation \citep[e.g.][]{helou85,dejong85,yun01,xu94,mauch07,Willott03}. Massive stars blow out ionized bubbles, generating radio free-free emission correlated with the ionizing flux. Some fraction of these ionizing photons are absorbed by dust clouds and reprocessed into far-infrared emission \citep{xu94}. At radio frequencies lower than $\sim10$\,GHz, synchrotron dominates over free-free emission, and the correlation is thought to arise from the acceleration of cosmic ray electrons in these stars' supernovae. 

Our approach is different. For all the advantages of broad band intensity mapping, sources cannot be localized to specific redshifts, meaning that it is foreground fluxes, not luminosities, whose correlations are of interest. Of course compact foregrounds may be masked or subtracted to some residual level, but any correlation of these residual foreground fluxes could bury the EOR correlation. We begin in this section by analyzing foreground fluxes as a function of masking depth, and in the next section turn to the foregrounds in residual images below the detection limit of these catalogs. 

\begin{figure*}[h]
\centering
\includegraphics[width=5.5in]{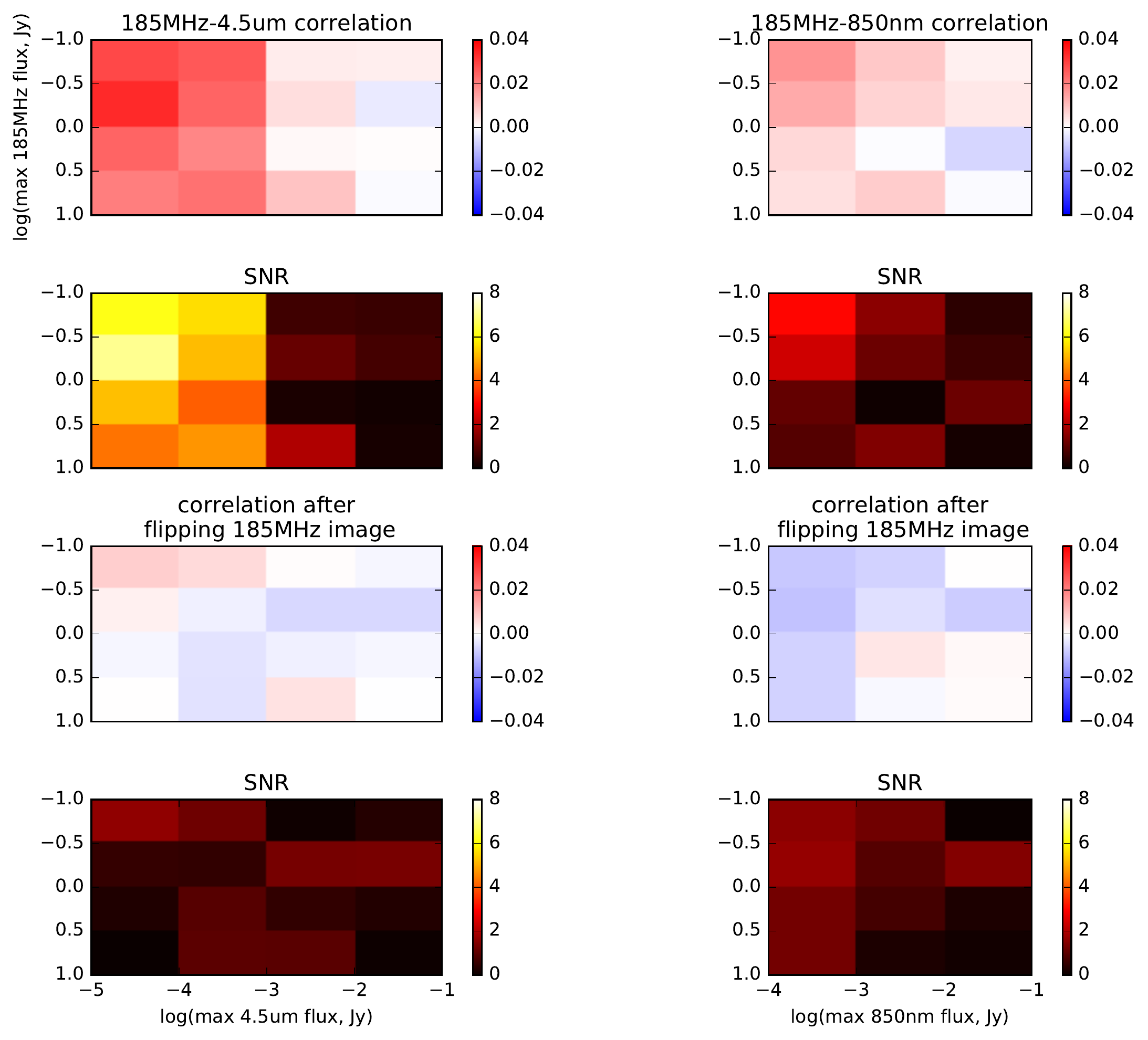}
\caption[Image space correlation coefficient between 185\,MHz and 4.5\,$\mu$m sources (left), and 185\,MHz and 850\,nm sources (right).]{Image space correlation coefficient between 185\,MHz and 4.5\,$\mu$m sources (top left), and 185\,MHz and 850\,nm sources (top right), both as a function of radio and infrared flux cuts. In the second row we calculate the SNR in each bin. The bottom two rows show the correlations and SNRs after flipping the 185\,MHz image about its vertical axis before the correlation calculation, giving an independent estimate of the noise. A significant correlation between 185\,MHz and 4.5\,$\mu$m sources appears after masking infrared sources down to 1\,mJy, which we show in Fig. \ref{fig:wisecolorcolor} cleanly removes the stars from the sample. The 185\,MHz and 850\,nm sources exhibit a marginal correlation after masking down to 1\,mJy, though it does not appear significant in comparison to the noise correlation seen after flipping the radio image. We show in Fig. \ref{fig:sdssstarsgals} that this is due to increased stellar contamination at 850\,nm than at 4.5\,$\mu$m.}
\label{fig:correlationsandSNRs}
\end{figure*}

We begin by gridding all three catalog fluxes in Jy to the $20^\circ\times20^\circ$ grid centered at (RA,Dec) = $(0, -30^\circ)$ depicted in Fig. \ref{fig:surveyoverview} at $5'$ resolution, and calculating the zero lag correlations between the images as
\begin{equation}
\label{eqn:imagecorrdef}
	c = \frac{\langle I_\rad I_\IR\rangle-\langle I_\rad\rangle\langle I_\IR\rangle}{\sqrt{(\langle I_\rad^2\rangle -\langle I_\rad\rangle^2)(\langle I_\IR^2\rangle -\langle I_\IR\rangle^2)}}
\end{equation}
where $\langle\rangle$ denotes an average over pixels, and the uncertainty due to sample variance is $\Delta c\approx N_\text{pix}^{-1/2}$, where $N_\text{pix}$ is the total number of pixels in the image. Between MWA and WISE catalogs we find $c=-0.003\pm0.005$ and between MWA and ATLAS catalogs we find $c=0.001\pm0.005$. Both are consistent with zero, as expected, as the brightest sources in both infrared catalogs are likely stars, whose radio emission is expected to be negligible. As a first experiment, we recalculate these correlations after excluding the brightest 10\% of sources in all three catalogs, effectively masking down to $10^{-3.75}$\,Jy at 4.5\,$\mu$m and $10^{-2}$\,Jy at 850\,nm, and find $c_\text{MWA--WISE}=0.031\pm0.005$ and $c_\text{MWA--ATLAS}=0.0086\pm0.005$. The former is a $6\sigma$ detection, and merits some investigation. How does this apparent correlation depend on the flux cut? What is it due to? And what does it mean for broad band correlation experiments? Further, does the MWA--ATLAS correlation remain consistent with zero at stricter flux cuts?

To begin to answer these questions, we plot in Fig. \ref{fig:correlationsandSNRs} the 185\,MHz--4.5\,$\mu$m correlation (top left) and 185\,MHz--850\,nm correlation (top right) as a function of the masking depth (i.e., the maximum flux of remaining sources). We plot the SNRs of these correlation measurements, taking the noise to be $N_\text{pix}^{-1/2}$ as described above, in the next row. Note that adjacent cells in the correlation matrix plots are somewhat correlated, so a consistent positive sign is not in and of itself evidence of significance. We assess significance by comparing of each correlation measurement individually with the expected noise (the SNR), as well as by checking that the correlation vanishes when the 185\,MHz image is flipped (bottom two rows). 

The 185\,MHz and 4.5\,$\mu$m catalogs exhibit a positive correlation peaking at $0.0332\pm0.005$ after masking infrared sources down to 10$^{-4}$\,Jy (18.9\,mag) and radio sources down to 1\,Jy, and remains significant down to the completeness limits of these catalogs. There is no significant correlation detection after flipping the 185\,MHz image, indicating this detection is not an artifact of the analysis, or of primary beam or vignetting effects. The 185\,MHz and 850\,nm catalogs exhibit a marginal $3\sigma$ correlation after masking infrared sources down to 10$^{-3}$\,Jy (16.4\,mag) and radio sources down to 0.3\,Jy, though it does not appear significant in comparison to the level of correlation noise in the flipped image.

\begin{figure*}[h]
\centering
\includegraphics[width=6in]{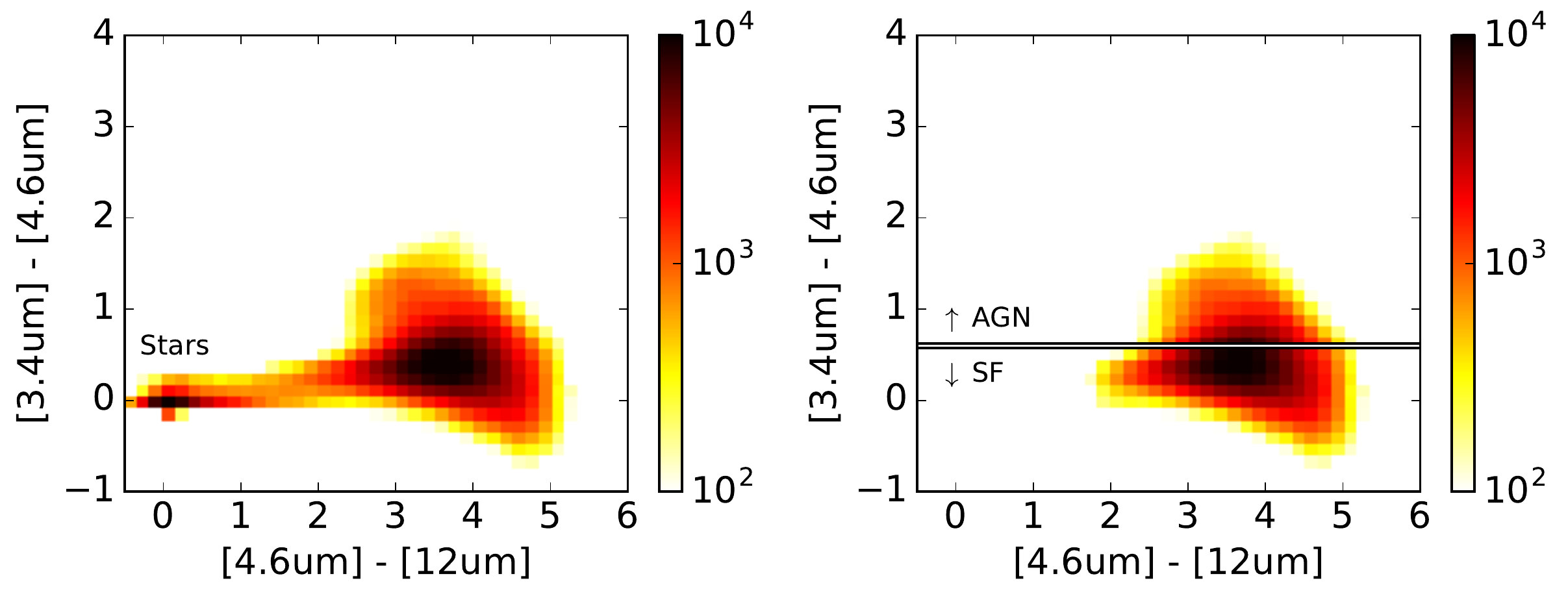}
\caption[Color-color plots of ALLWISE sources showing that cutting out the brightest 10\% cleanly removes the stars.]{The ALLWISE sources are plotted in the color--color space of \citet{Wright2010} prior to any flux cuts (left panel), showing cluster of stars near (0,0). The colorbar shows the number of sources in each cell. Then after cutting out the brightest 10\% of sources (fainter than 18.25\, AB mag), the stars are cleanly removed. We roughly split the remaining sources into AGNs ($W_{12}>0.6$) and starburst galaxies (SF) ($W_{12}<0.6$) \citep{mingo16}, where  $W_{12}\equiv$ [3.4\,$\mu$m] -- [4.6\,$\mu$m].}.
\label{fig:wisecolorcolor}
\end{figure*}

To understand these findings, we begin by investigating which 4.5\,$\mu$m sources are responsible for this correlation. We select the subset of sources detected in the WISE 3.4\,$\mu$m, 4.5\,$\mu$m, and 12\,$\mu$m bands and plot them (Fig. \ref{fig:wisecolorcolor}, left panel) in the $W_{23}\equiv$ [4.6\,$\mu$m] -- [12\,$\mu$m] versus $W_{12}\equiv$ [3.4\,$\mu$m] -- [4.6\,$\mu$m] color-color space used by \citet{Wright2010} to illustrate the separation between different types of sources. \citet{nikutta14} study more quantitatively how sources separate in this space, finding that stars are isolated in the region $W_{12}=-0.04\pm0.03$, $W_{23}=0.05\pm0.04$ (1$\sigma$). In the right panel, we plot the faintest 90\% of sources (fainter than 18.25 mags at 4.6\,$\mu$m) in the same color-color space and observe this cut effectively cleanly excludes nearly all the stars. This explains the detection of a 185\,MHz--4.5\,$\mu$m correlation only after masking the brightest 10\% of sources. 

\begin{figure*}[h]
\centering
\includegraphics[width=7in]{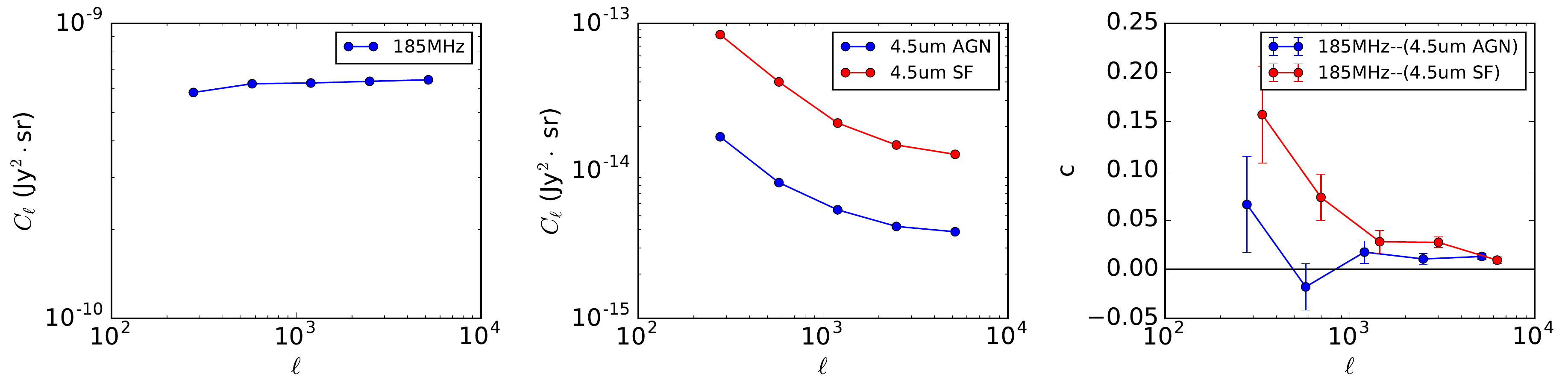}
\caption[Auto and cross spectra of 185\,MHz and 4.5\,$\mu$m sources, for both AGN and SF cuts of the infrared catalog.]{Power spectrum of 185\,MHz sources (left panel), 4.5\,$\mu$m sources (center panel), and coherence between 185\,MHz and 4.5\,$\mu$m sources (right panel). We roughly separate the 4.5\,$\mu$m sources into starforming galaxies (SF) and active galactic nuclei (AGN) as illustrated in Fig.  \ref{fig:wisecolorcolor}. We find that the 185\,MHz--4.5\,$\mu$m correlation observed in Fig. \ref{fig:correlationsandSNRs} holds only for the starforming galaxies in the infrared sample. }
\label{fig:wisexspec}
\end{figure*}

To further probe which mid infrared sources are responsible for this correlation, we make a rough cut to separate quasars and active galactic nuclei (AGN) ($W_{12}>0.6$) from starforming galaxies (SF) ($W_{12}<0.6$) \citep{mingo16}. In Fig. \ref{fig:wisexspec} we plot the power spectrum of 185\,MHz sources (left panel), 4.5\,$\mu$m AGN (center panel, blue points), and 4.5\,$\mu$m starforming galaxies (center panel, red points). In the right panel we plot the coherence (i.e., normalized cross spectrum) of the 185\,MHz catalog with AGN (blue) and with starforming galaxies (red). The AGN cut exhibits no significant correlation with the 185\,MHz sources, while the starforming galaxy cut exhibits a significant correlation rising from a few percent at $\ell\sim7000$ to 16\% at $\ell\sim300$. 

The fall of the correlation towards high $\ell$ is likely due to the MWA's $3'$ resolution at 185\,MHz, corresponding to a maximum $\ell$ of roughly 4000. Both the falling 4.5\,$\mu$m catalog power spectrum and the relatively flat 185\,MHz power spectrum are functions of the detailed properties of these surveys. \citet{tegmark02,dodelson02} show that the galaxy angular power spectrum $C(\ell)$ is approximately equal to the 3D matter power spectrum $P(k(\ell))$ convolved with with a window function which depends on the redshift coverage and flux limit of the sample. The matter power spectrum is known to rise as $k^{1}$ for $k\lesssim0.02$\,h/Mpc before falling as $k^{-3}$. Galaxy surveys typically probe the regime just after the turnover where the slope is transitioning from 0 to -3 \citep{tegmark02b}. In order to maximize its sensitivity to low surface EOR 21\,cm emission, the MWA was designed as a relatively compact array in comparison to higher resolution radio interferometers such as the Very Large Array. This low resolution makes the MWA catalog severely flux limited \citep{PattiCatalog1}, which in turn effectively masks many galaxies which would otherwise be seen. This large masked volume translates into a wide Fourier convolution kernel, explaining why the MWA catalog power spectrum is so flat. 

Lastly, we hypothesize that the absence of an observed 185\,MHz--850\,nm correlation is due to the larger fraction of stars in 850\,nm images than in 4.5\,$\mu$m images. To check this, we study the fraction of stars and galaxies in a similar Galactic field observed by Sloan Digital Sky Survey (SDSS) \citep{sdssiii} DR13 \citep{sdssdr13}. SDSS catalog\footnote{\ULurl{http://skyserver.sdss.org/dr13/en/tools/search/rect.aspx}} objects are classified as either stars or galaxies using \textit{ugriz} photometry, though unfortunately SDSS doesn't reach the declination of our MWA field at (RA,Dec)=($0,-27^\circ$). We thus use a $5^\circ$ field centered at (RA,Dec)=($205^\circ$,$22^\circ$), which has the same galactic longitude but is flipped to the other side of the galactic plane, giving a similar line of sight through the galaxy.

\begin{figure*}[h]
\centering
\includegraphics[width=3in]{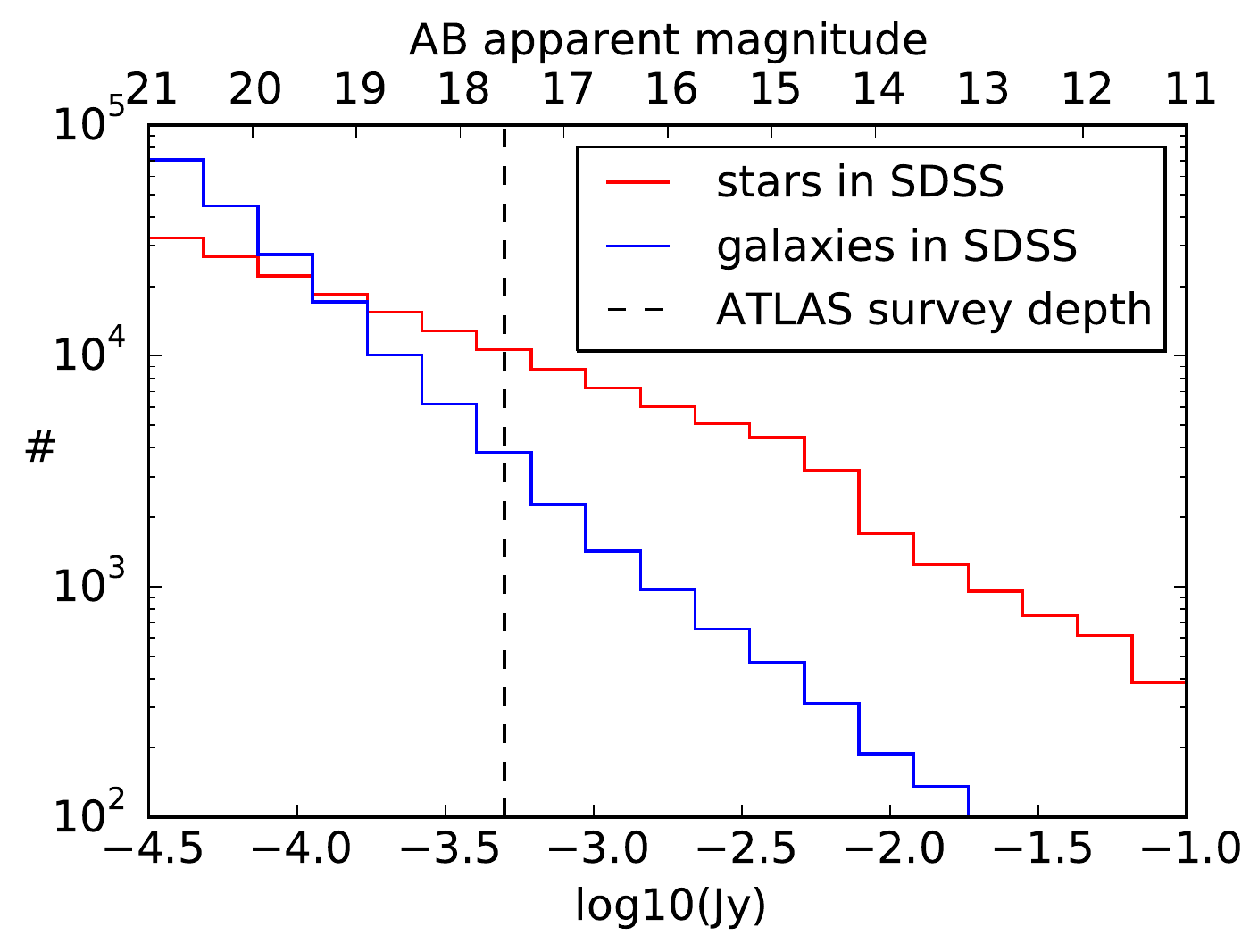}
\caption[Histogram of SDSS fluxes of stars and galaxies in a similar field to our ATLAS field.]{Histogram of fluxes of stars and galaxies in a $5^\circ$ SDSS field at the same galactic longitude as the MWA field, but flipped to the other side of the galactic plane. These fluxes are in the SDSS i band $(762\pm65)$\,nm. }
\label{fig:sdssstarsgals}
\end{figure*}

In Fig. \ref{fig:sdssstarsgals} we plot a histogram of the fluxes of the objects marked as stars (red) and galaxies (blue) in this field at the SDSS i band $(762\pm65)$\,nm. We observe that stars dominate the field above  $10^{-4}$\,Jy, this is a factor of a few below the survey depth of our wide ATLAS 850\,nm image, and thus none of the flux cuts explored above were deep enough to reveal the extragalactic sources. This is consistent with the fact that the 185\,MHz--4.6\,$\mu$m correlation only appeared after the stars were removed, a procedure easier in the mid-infrared than the near-infrared. 

\subsection{Simulations of distance-induced flux correlation}
\label{sec:fluxcorsims}
Let us now consider why the 4.5\,$\mu$m SF sample is 5--15\% correlated with the 185\,MHz catalog, while the AGN sample is not. Of course some slight correlation is expected simply because brighter AGN typically reside in more massive galaxies, which are typically brighter in stars (see, for instance, Fig. 1 in \citep{seymour07} or Fig. 4 in \citep{Willott03}), but \citet{mauch07} find no strong correlation between radio and \textit{near}-infrared luminosities. As discussed above, though, broadband correlation intensity mapping experiments are affected not only by luminosity correlations but by flux correlations as well. We show in this section that fluxes in two different bands may appear correlated due to geometric effects even when their intrinsic luminosities are completely independent of each other. By geometric effects we refer to to the fact that more distant objects are generally weaker in all bands than nearer objects. 

We first make a few approximations to build intuition, then simulate the effect as a function of source masking depth. Consider a sky survey with fixed field of view of a set of objects with uncorrelated infrared and radio emission. By uncorrelated we mean that the infrared and radio luminosities are independent random variables determined by the infrared and radio luminosity functions, respectively. Assume that the objects are uniformly distributed in space out to $z\sim0.5$, and work in Cartesian space for simplicity. We are interested in the effective correlation between radio and infrared fluxes in the same sky pixels, but let us approximate this by calculating the correlation between source fluxes in the two bands. Starting from Eqn. \ref{eqn:imagecorrdef}, we have
\begin{equation} 
	c = \frac{\langle F_\rad F_\ir \rangle-\langle F_\rad\rangle\langle F_\ir\rangle}{\sqrt{(\langle F_\rad^2\rangle-\langle F_\rad\rangle^2)(\langle F_\ir^2\rangle-\langle F_\ir\rangle^2)}}
\end{equation}
where $\langle\rangle$ denote an average over sources in the catalog. Then making the approximation that the radio and infrared luminosity functions are independent of line of sight distance, we can rewrite this equation in terms of moments of these luminosity functions and the distribution of object distances using $F_i=L_i/4\pi D^2$ for $i=$radio,IR.
\begin{equation}
\label{eqn:cresult}
	c = \frac{\beta-1}{\sqrt{(\beta\alpha_\rad-1)(\beta\alpha_\ir-1)}}\approx\frac{1}{\sqrt{\alpha_\rad \alpha_\ir}}
\end{equation}
where $\beta\equiv\langle D^{-4}\rangle/\langle D^{-2}\rangle^2$ and $\alpha_i=\langle L_i^2\rangle/\langle L_i\rangle^2$, and again $\langle\rangle$ denote an average over sources in the catalog. The approximation in the last part of this equation results from the fact that for typical parameters and luminosity functions (see below) $\beta>>1$ and $\alpha>1$. 

For a survey of a fixed angular field of view, uniform spatial distribution of objects, and Cartesian spacetime, the distribution of object distances is $\rho(D)=\rho_0 D^2$. Assuming all the objects are between distances $D_\mintext$ and $D_\maxtext$, the normalization constant is $\rho_0=3/(D_\maxtext^3-D_\mintext^3)$.
\begin{eqnarray}
	\beta &=& \frac{\int_{D_\mintext}^{D_\maxtext}D^{-4}\rho(D)dD}{\left(\int_{D_\mintext}^{D_\maxtext}D^{-2}\rho(D)dD\right)^2}\\
	 &=&\frac{D_\maxtext^2+D_\maxtext D_\mintext+D_\mintext^2}{3D_\maxtext D_\mintext}
\end{eqnarray}

We observe that the radio--infrared flux correlation $c$ for some type of objects is a function of their radio and infrared luminosity functions. In fact, we can see immediately that if the luminosity distributions are wide, their $\alpha$'s are large, and $c$ is small. Conversely, if the luminosity functions are narrow, then the distance to the sources plays a more significant role in determining their fluxes, and so $c$ is larger. 

\begin{figure*}[h]
\centering
\includegraphics[width=6in]{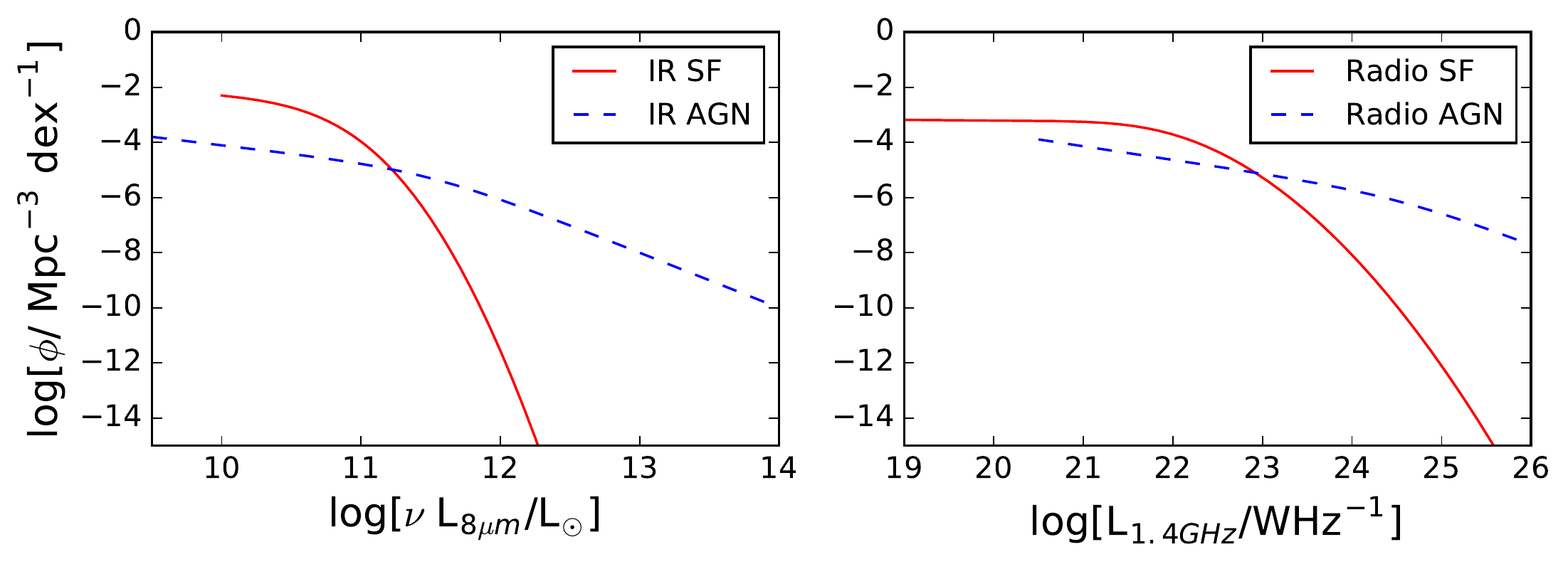}
\caption[AGN and SF luminosity functions at 8\,$\mu$m from (left) and at 1.4\,GHz (right).]{AGN and SF luminosity functions at 8\,$\mu$m from \citet{fu10} (left panel) and at 1.4\,GHz from \citet{mauch07} (right panel).}
\label{fig:luminosityfunctions}
\end{figure*}

To quantify whether this effect can explain our measured radio--infrared correlation in SF galaxies and the lack of one in AGN, we use AGN and SF luminosity functions at 1.4\,GHz from \citet{mauch07} (Fig. \ref{fig:luminosityfunctions}, right panel) and at 8\,$\mu$m from \citet{fu10} (left panel). The former describe galaxies at $z<0.3$, while the latter describe galaxies at $z\sim0.6$. In principle we should use luminosity functions at our actual radio and infrared bands, 185\,MHz and 4.5\,$\mu$m, and of course this analysis could be extended using proper redshift-dependent luminosity functions, though we find that our simplified analysis suffices to explain our earlier correlation measurements. We leave a more detailed study for future work. Indeed \citet{prescott16} find that the AGN and starforming galaxy radio luminosity functions at lower frequencies, specifically at 325\,MHz, closely follow those at 1.4\,GHz up to an overall scaling which cancels out of our correlation coefficient. We use approximately the same range of luminosities used by \citet{mauch07} and \citet{fu10}, and adjust the minimum luminosities slightly to achieve the same number density of each type of object in both radio and infrared surveys. Above these limiting magnitudes (see Fig. \ref{fig:luminosityfunctions}), the number density of AGN is $\sim0.0020$/Mpc$^3$ and that of SF is $\sim0.00011$/Mpc$^3$. In the end we find that our results are only weakly sensitive to these luminosity minima as their faint ends become less and less significant in real, flux-limited samples. 

We pick fiducial survey parameters of $d_\mintext=20$\,Mpc and $d_\maxtext=3000\,Mpc$ ($z_\maxtext=0.75$), giving $\beta\approx50$.  Using the above luminosity functions, we find $\alpha_{\text{SF,IR}}=1.474$, $\alpha_{\text{SF,rad}}=14.56$, $\alpha_{\text{AGN,IR}}=22.97$, $\alpha_{\text{AGN,rad}}=257.5$. These values agree with qualitative observation that the AGN luminosity function is wider than the SF luminosity function in both radio and infrared bands (Fig. \ref{fig:luminosityfunctions}). These values give a predicted radio--infrared correlation of 0.21 for SF and 0.01 for AGN agreeing with our finding of a significant radio--infrared correlation for SF and near-zero correlation for AGN. The exact values deviate from our measurements for a number of reasons. The MWA and WISE catalogs are not matched in depth or redshift coverage, and thus don't survey an exactly overlapping set of radio and infrared sources. Further, these calculations assume a volume limited survey, in contrast to our flux limited radio and IR surveys. Additionally, real world luminosity functions can exhibit redshift evolution. Of course, our measurement in Sec. \ref{sec:catcorrelations} did not even split up the MWA catalog into separate AGN and SF subsets as such detailed characterization of low frequency radio foregrounds remains an active area of research. Lastly, the limited MWA resolution pushes the observed correlation with infrared images to zero at high-$\ell$, suppressing the overall correlation computed in image space. Future work will needed to quantify these effects in greater detail and assess their significance in EOR cross spectrum measurements.

\begin{figure*}[h]
\centering
\includegraphics[width=6in]{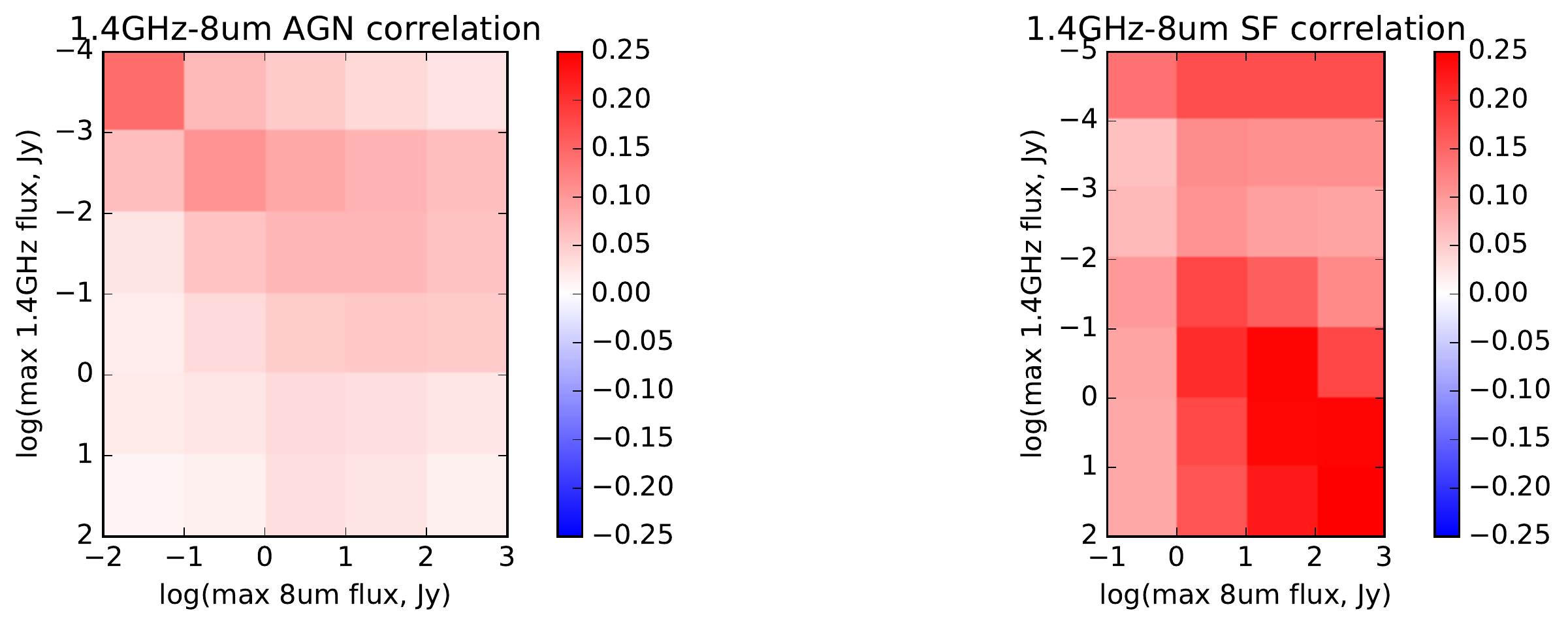}
\caption[Apparent correlation of radio and mid-infrared fluxes from a mock survey.]{Apparent correlation of radio and mid-infrared fluxes from a mock volume-limited survey with independent radio and mid-infrared luminosities using the luminosity functions from Fig. \ref{fig:luminosityfunctions} for AGN and SF. Without any flux cuts (lower right corner of each plot) we observe a significant correlation between radio and mid-infrared fluxes for SF objects, and a near-zero correlation for AGN. This agrees with our measurements on real 185\,MHz and 4.5\,$\mu$m sources in Fig. \ref{fig:correlationsandSNRs}. As fainter and fainter objects are masked, we observe that the AGN correlation gradually strengthens, while the SF correlation weakens somewhat. }
\label{fig:simagnlfcorrelations}
\end{figure*}

Can this unwanted radio--infrared foreground correlation be mitigated by masking the brightest sources? Using the luminosity functions presented above, we simulate radio and infrared surveys for each of AGN and SF. We begin by generating the mock radio catalogs of AGN and SF, choosing a Poisson random number of each in each of 400 logarithmic luminosity bins. Using logarithmic bins best samples the large dynamic range of the luminosity functions. We distribute the objects uniformly over a volume $D_\maxtext=cz_\maxtext/H_0=3212$\,Mpc deep and $\theta_{\text{FOV}}D_\maxtext=1121$\,Mpc wide, then pick a random infrared luminosity for each radio object from the appropriate infrared luminosity function. Finally we plot in Fig. \ref{fig:simagnlfcorrelations}, along the lines of Fig. \ref{fig:correlationsandSNRs}, the predicted 1.4\,GHz--8\,$\mu$m correlation of our mock AGN and SF catalogs after masking down to a maximum radio and infrared flux. As we saw above, without any flux cut we find a roughly 20\% radio--infrared correlation for SF and negligible correlation for AGN. As we mask fainter and fainter sources, the AGN correlation generally increases to the 5--10\% level, while the SF correlation first increases, then decreases after masking down to 10$^{-4}$\,Jy. With increasing mask depth, these correlations do not necessarily approach zero monotonically, and more detailed modeling of effective foreground flux correlations will be necessary in real world intensity mapping correlation experiments probing the EOR. In the next section we move beyond the bright sources and study the magnitudes and correlation properties of the residual radio and infrared foregrounds in our MWA and ATLAS observations.

\section{Residual foregrounds and cross spectrum limits}

In this section we characterize the power spectra and correlation properties of the residual 185\,MHz and 850\,nm foregrounds after subtracting and masking  the the bright sources identified by the surveys discussed in the previous section.

\subsection{Residual 21\,cm foregrounds}
\label{sec:res21fgs}

We begin by quantifying the 185\,MHz foreground residuals in angular power spectrum measurements. In broad band (i.e., multifrequency synthesis) images, thermal noise quickly integrates below foreground residuals. Reaching the cosmological signal should therefore be a matter of foreground mitigation and not the long time averages needed to measure the 3D power spectrum \citep[e.g.][]{beardsley13,PoberNextGen}. We check this hypothesis, asking how how much observation time is required to achieve the best foreground subtraction.

Working with the MWA image products presented in Sec. \ref{sec:mwaobservations}, we compute the angular power spectrum as
\begin{equation}
	C(\ell)=\frac{\sum_{\vec{\ell}}|\tilde{I}_\text{uni}(\vec{\ell})\tilde{I}_w(\vec{\ell})|^2}{\sum_{\vec{\ell}}|\tilde{I}_w(\vec{\ell})|^2}\frac{d\theta^2}{N^2}
\end{equation}
summing over all the $\vec{\ell}$ values in each $\ell$ bin, where $N$ is the number of pixels of size $d\theta$ on each side of the square image. Note that $\ell=2\pi u$, where $u$ is the Fourier dual to angle from the field center assuming the small angle approximation. We estimate the thermal noise power spectrum by computing the power spectrum of the difference between the interleaved odd and even cubes discussed earlier, which contains only thermal noise.

\begin{figure}[h]
\centering
\includegraphics[width=3.5in]{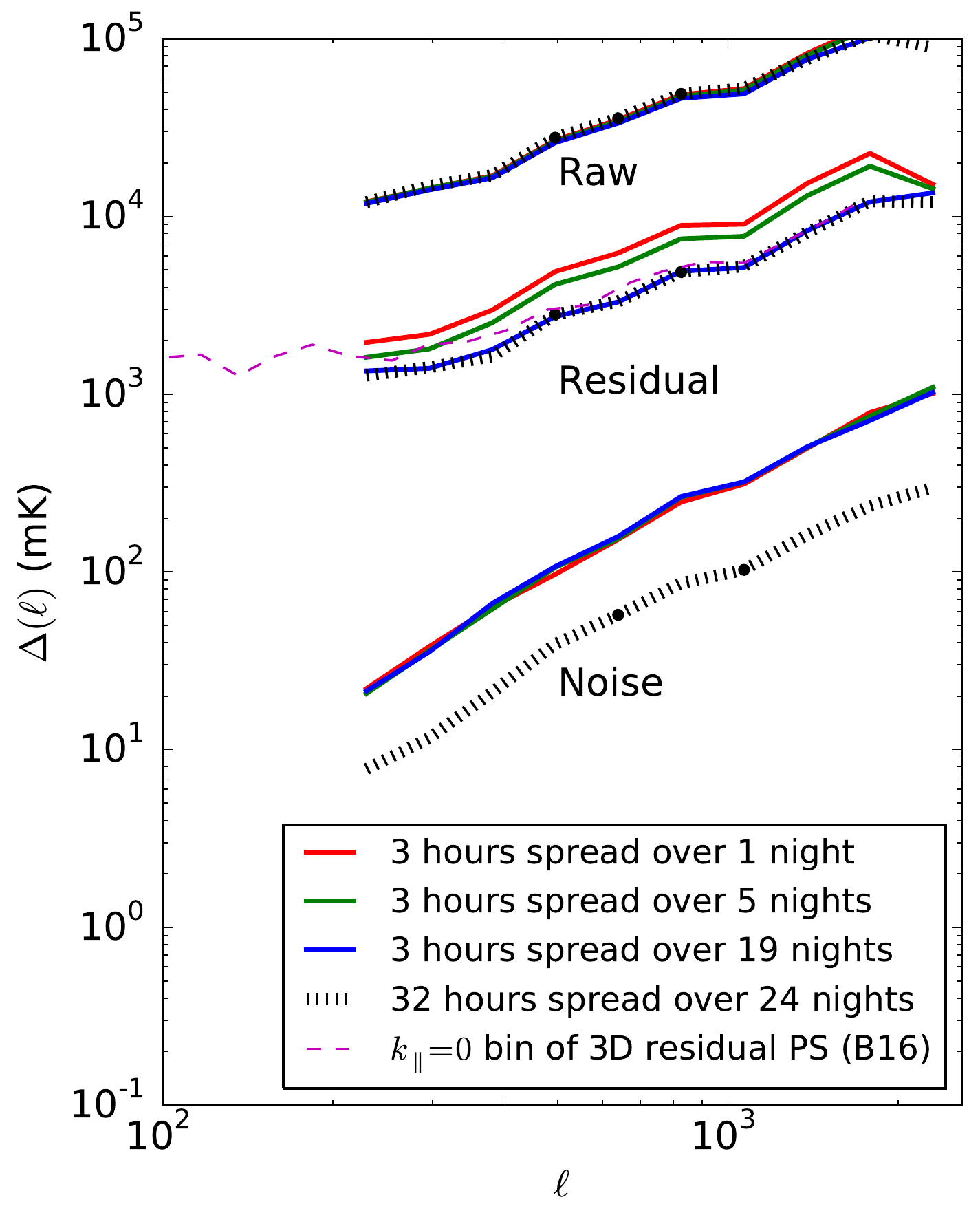}
\caption[Raw, residual (post foreground subtraction), and noise power spectra of 185\,MHz broad band images from various data selections.]{Raw, residual (post foreground subtraction), and noise power spectra of 185\,MHz broad band images from various 3\,hour data selections (solid lines) and from the 32\,hour data selection used by  \citet{beardsley16} (black dashed). The $uv$ plane is nearly filled after only 3\,hours, yet we find that spacing these $\sim100$ two-minute integrations over many independent nights reduces the foreground residuals by a factor of up to $\sim4$ in power. As a cross check on our analysis, we plot the $k_\parallel=0$ bin of the 3D power spectrum of \citet{beardsley16} converted to an angular power spectrum using Eqn. \ref{eqn:convertP3toP2} (magenta dashed).}
\label{fig:respspecspacingsstudy}
\end{figure}

We plot in Fig. \ref{fig:respspecspacingsstudy} the power spectra of 185\,MHz broad band images from  3\,hour data selections, spread over 1 (red solid), 5 (green solid), and 19 (blue solid) nights. We also make a broad band image of the same 32\,hour data set used by \citet{beardsley16} (black dashed). We plot the power spectra of the raw (pre foreground subtraction), residual (post foreground subtraction), and noise (difference between successive integrations) images out to $\ell=2600$, corresponding to a maximum baseline length of $\sim$700\,m. Beyond this, the $uv$ coverage becomes sparse, introducing artifacts in the application of gridding and uniform weighting, though a more sophisticated analysis could likely use these longer baselines. We cross-check our imaging and power spectrum analysis by comparing the  $k_\parallel=0$ bin of the 3D power spectrum of \citet{beardsley16} converted to an angular power spectrum using Eqn. \ref{eqn:convertP3toP2} (magenta dashed). 

We find that the raw power spectra of the different 3\,hour data sets agree with each other and with that of the 32\,hour data set, as expected given that foregrounds overwhelm thermal noise in a broad band image. Interestingly, the residual power spectrum decreases as the 3 hours are spread over more and more nights until it reaches the level of the deep 32 hour integration, a factor of $\sim4$ lower in power than in the single night analysis. These findings could be explained by slight ionosphere-related errors which limit the accuracy of each night's calibration, and thus, of its foreground subtraction. Further work would be needed to understand this effect in detail, but for now our conclusion is that the 32\,hour integration has the best foreground subtraction, and we use this deep cube in our later analyses. Note that as expected, the thermal noise of the deep integration is 10 times lower in power than the 3\,hour integrations. Because these are band averaged images, even the 3\,hour thermal noise is at least 100 times lower than its foreground residuals.

\subsection{Residual IR foregrounds}
\label{sec:resirfg}

We proceed to generate foreground-masked 850\,nm images of each of the four deep ATLAS integration fields shown in Fig. \ref{fig:surveyoverview}. Each of these fields is a stack of 9 30\,sec exposures with 5$^\circ$ field of view, dithered such that the overlap is a $\sim4^\circ$ region. By confining ourselves to this overlap region, we avoid the background discontinuities which affect many nominally wide field infrared image datasets whose mosaicing introduces significant background patchiness. \citet{mw15} have demonstrated that complex fitting \citep{fixen00} can help reduce such patchiness in small mosaics ($\sim10$ arcminutes), but further work is required to study whether such techniques can be applied to much larger fields.

\begin{figure*}[h]
\centering
\includegraphics[width=7in]{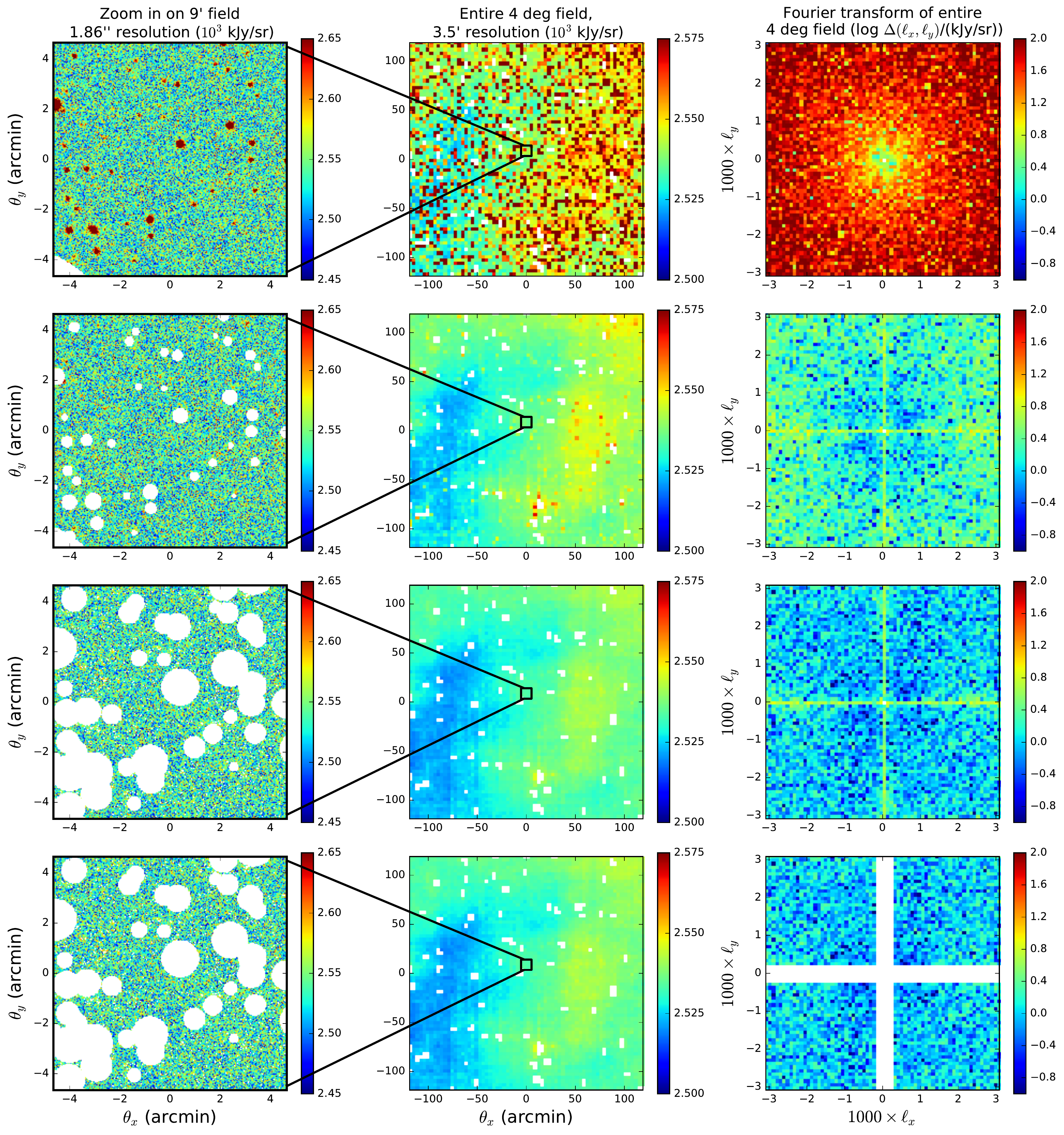}
\caption[Images of our 850\,nm field after various stages of foreground masking.]{The rows of this figure illustrate our four stages of foreground removal on ATLAS images, as detailed in Sec. \ref{sec:resirfg}. We apply this process to all four $4^\circ$ deep ATLAS fields shown in Fig. \ref{fig:surveyoverview}, but here show the one centered at (RA,Dec)$=2.74^\circ, -24.79^\circ$) for illustration. The first row shows the results of masking $4'$ around all nearly saturated regions; the second row shows the results of masking out to $5\sigma$; the third row shows the results after masking out to $12\sigma$ and all pixels above 70\,kJy/sr over the background, and the last row shows the results after masking the nearly horizontal or vertical Fourier modes. In each row, we show the central $9'$ field at $1.86''$ resolution (left), the entire $4^\circ$ field after coarse gridding to $3.5'$ (center), and the FFT of the coarse gridded field to highlight systematics (right). Note that the left and center panels of the bottom row are identical to those in the row above it, illustrating that the image space mask is the same, only the Fourier mask is different.}
\label{fig:bigfgmaskingstudy}
\end{figure*}

Our approach is to mask each image at ATLAS's native $1.86''$ resolution, then coarse grid down to $3.5'$, approximately the resolution of the MWA, taking each coarse pixel's value as the average of all unmasked fine pixels within it. If fewer than 10\% of its fine pixels remain unmasked, we consider the whole coarse pixel masked to avoid introducing too much noise variation between different coarse pixels. For illustrative purposes, we proceed in the following four stages, which we illustrate in Fig. \ref{fig:bigfgmaskingstudy} for the ATLAS field centered at (RA,Dec)$=(2.74^\circ, -24.79^\circ$) (the top right red box in Fig. \ref{fig:surveyoverview}). Each row shows the result of an additional masking stage, as outlined below. The left column shows a typical $9'$ field to illustrate the masking up close; the center column shows the resulting coarse binned image with $3.5'$ resolution; and the right column shows the FFT of the center image (plotted as $\log[\Delta(\ell)/(\text{kJy/sr})]$) in order to identify detector systematics. 

\begin{enumerate}
	\item \textbf{Mask saturated regions.} (Fig. \ref{fig:bigfgmaskingstudy}, row 1) Nearly saturated pixels are associated with nearby bright stars which would dominate fluctuation measurements, we thus mask $4'$ around all pixels within 30\% of saturation (white regions in left and center columns). This wide mask removes the broad wings of the PSF revealed by these extremely bright sources. Roughly 92\% of fine pixels remain unmasked after this step, and 96\% of coarse pixels remain.
	\item \textbf{Mask sources to 5$\sigma$.} (Fig. \ref{fig:bigfgmaskingstudy}, row 2) We mask circular regions with radius equal to five times the profile RMS along the minor axis of each source as measured by SExtractor (see Sec. \ref{sec:catalogs}). The reason we use the minor axis RMS is that vertical charge leakage in the CCD array results in unrealistically large major axis RMS measurements for bright sources. After this stage $\sim87$\% of fine pixels remain unmasked, and the fraction of coarse pixels remaining is unaffected.
	\item \textbf{Mask sources to 12$\sigma$ and mask other emission above 70\,kJy/sr over the background.} (Fig. \ref{fig:bigfgmaskingstudy}, row 3) We use a larger masking radius to remove the PSF wings around the 5$\sigma$ source masks. We also determine that the vertical CCD charge leakage can be isolated by looking for emission brighter than 70\,kJy/sr over the background, and that it can be flagged without cutting into the shot noise between sources. 58\% of fine pixels remain unmasked after this step, and again the fraction of coarse pixels remaining is unaffected.
	\item \textbf{Mask horizontal and vertical Fourier modes.} (Fig. \ref{fig:bigfgmaskingstudy}, row 4) The previous two stages revealed detector artifacts within $\Delta\ell \sim100$ of $\ell_x=0$ and $\ell_y=0$. These compact Fourier systematics correspond to slight horizontal and vertical discontinuities in the center image due to imperfect gain matching between 16 different amplifiers which process different rectangular regions of the CCD array. We conservatively mask Fourier modes with $|\ell_x|<200$ or $|\ell_y|<200$ to eliminate this effect.
\end{enumerate}

Note that these 850\,nm deep observations were recorded during near new moon conditions, and we find that the mean air glow in source-free regions is $\sim3\times10^3$ kJy/sr, of order 19 AB mag/arcsec$^2$. For comparison, \citet{sullivan12} measure a 1020\,nm continuum air glow brightness (i.e., after spectrally masking the OH lines) of $20\pm0.5$ AB mag/asec$^2$ far away from the moon. 

We proceed to characterize the residual infrared fluctuations in power spectrum space, using the optimal quadratic estimator to properly account for the masking of coarse pixels. This estimator was introduced to astronomy by \citet{Maxpowerspeclossless} to recover CMB power spectra from maps with arbitrary survey geometries and noise properties, and has recently been revived for 3D power spectrum analysis of 21\,cm data by \citet{X13, dillonneben, LT11, DillonFast, ali15}. We employ this estimator in a manner more similar to the original CMB case, using it to account for pixel masking in broad band power spectrum estimation. We briefly summarize the estimator here, and refer to \citet{X13} for a more detailed description.

We label the normalized estimator of the power in bin $\alpha$ as $p_\alpha$, related to the unnormalized estimator $q_\alpha$ as $\mathbf{p} = \Mb \mathbf{q}$. Bold lower-case letters are vectors, while bold upper-case letters are matrices. The unnormalized estimator is given by
\begin{equation}
q_\alpha = \frac{1}{2}(\xb-\langle\xb\rangle)^t \Cb^{-1} \Cb_{,\alpha}\Cb^{-1}(\xb-\langle\xb\rangle)
\end{equation}
where $\xb$ is a column vector containing all $N_\text{pix}$ pixel measurements, $\Cb$ is the pixel-pixel covariance matrix, and $\Cb_{,\alpha}\equiv d\Cb/dp_\alpha$ is the derivative of the covariance with respect to the power in bin $\alpha$. Note that $\Cb_{,\alpha} = \Ab^\dagger\Ab$, where $\Ab$ is a $N_\alpha\times N_\text{pix}$ with elements $A_{ij}=\exp(i\vec{\theta}_j\cdot\vec{\ell}_i)$. Here $\vec{\ell}_i$ refers to the $i$'th out of $N_\alpha$ $\vec{\ell}$ modes in bin $\alpha$. Note that $^t$ denotes a transpose, and $^\dagger$ donates a conjugate transpose.

The matrix of window functions (i.e., horizontal error bars) of the band powers $p_\alpha$, defined such that $\pb_\text{estimated}=\Wb\pb_\text{true}$ is given by $\Wb=\Mb\Fb$, where $\Fb$ is the Fisher matrix and $\Mb$ is an arbitrary invertible normalization function encoding the compromise between horizontal and vertical error bars. The covariance between the measured $p_\alpha$ values is given by $\mathbf{\Sigma} = \Mb\Fb\Mb^t$. \citet{X13} argue that taking $\Mb\propto \Fb^{-1/2}$ is a good compromise between small horizontal error bars and small vertical error bars. For simplicity, we take $\Mb=\Fb^{-1/2}$, and correct the normalization at the end by dividing each element of $\pb$ by the peak of the appropriate row of $\Wb$. In this case, the bandpower variances are given by the reciprocals of the peaks of the window functions used to normalize the bandpowers. We find that using the sum instead of the peak significantly biases the recovered bandpowers downward when the power spectrum is non-flat, as in our case. Lastly, the elements of the Fisher matrix are given by
\begin{equation}
\Fb_{\alpha\beta}=\frac{1}{2}\text{tr}\left(\Cb^{-1} \Cb_{,\alpha} \Cb^{-1} \Cb_{,\beta} \right)	
\end{equation}

Now we turn to application of this formalism to power spectrum estimation from our masked IR images. Later we will adapt it to estimation of the 21\,cm--IR cross spectrum. We take $\xb$ to be a vector of all IR coarse ($3.5'$) pixel values, with masked pixels set to zero. After gridding the high resolution images to reach this resolution, photon shot noise is negligible, and the image space covariance is the sum of the sample variance $\Cb_\text{signal}$ and the masking covariance $\Cb_\text{mask}$. $\Cb_\text{mask}$ is a diagonal matrix with $\infty$ for masked pixels, and 0 otherwise. In practice, we replace $\infty$ with a number $10^7$ times larger than the largest eigenvalue of $\Cb_\text{signal}$, finding that the results are not sensitive to this parameter. The sample variance is easily obtained by writing it in Fourier space, $\Cb_\text{ft}$, where it is a diagonal matrix with a guess of the true power spectrum on the diagonal, then Fourier transforming it into image space with Fourier transform matrix $\mathcal{F}$. Putting these together gives
\begin{equation}
\label{eqn:covFTwithmask}
	\Cb = \mathcal{F}^\dagger\Cb_\text{ft}\mathcal{F}+\Cb_\text{mask}
\end{equation}
Note that $\mathcal{F}$ is an $N_\text{pix}\times N_\text{pix}$ matrix with elements $\mathcal{F}_{ij}=\exp(-i \vec{\theta}_i\cdot\vec{\ell}_j)$, where $i$ runs over all pixels and $j$ runs over all Fourier cells. Said differently, a guess of the power spectrum is necessary to optimally downweight the sample variance noise on the estimated power spectrum. If the accuracy of the guess were in question, one could always iterate by feeding the estimated power spectrum back into the quadratic estimator, though in practice we find this is not necessary in this work.

\begin{figure*}[h]
\centering
\includegraphics[width=7in]{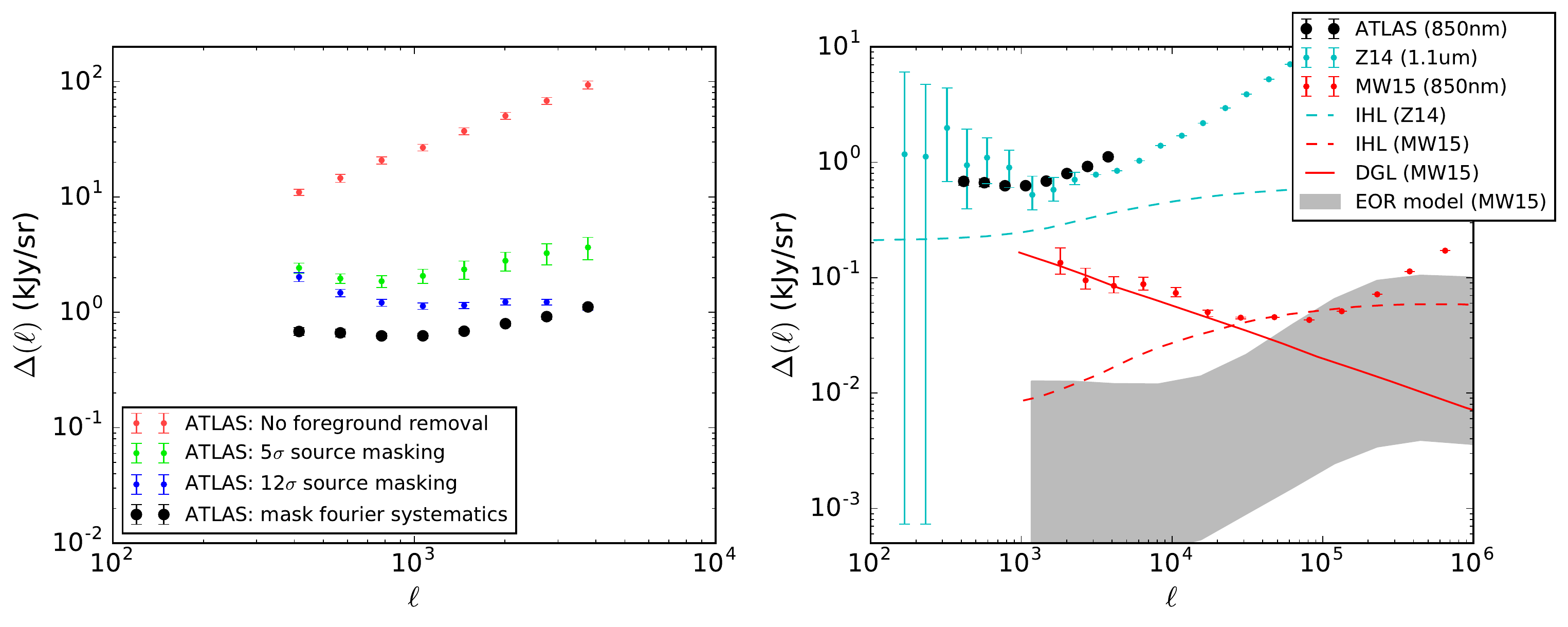}
\caption[Power spectra of our 850\,nm field after various stages of foreground masking.]{\textbf{Left panel:} Power spectra of our broad band ATLAS images after various stages of foreground removal as described in Sec. \ref{sec:resirfg}. The error bars show the sample variance noise estimated by the standard deviation over our four $4^\circ$ fields (see Fig. \ref{fig:surveyoverview}). Black dots show our final ATLAS 850\,nm power spectrum after all stages of foreground removal. \textbf{Right panel:} We compare our ATLAS power spectrum (black dots) to the 1.1\,$\mu$m power spectrum of \citet{zemcov14} (Z14) (cyan dots) using the CIBER experiment and the 850\,nm power spectrum of \citet{mw15} (MW15) using $10'$ Hubble mosaics. We show the intrahalo light (IHL) models of ZI4 (cyan dashed line) and MW15 (red dashed line), as well as the diffuse galactic light (DGL) model of the latter authors. These results show that finer angular resolution can help reduce the foregrounds in the $\ell\sim10^2-10^4$ modes that can be cross correlated with 21\,cm EOR intensity mapping experiments. The gray area shows the predicted EOR contribution to the infrared anisotropies from MW15.}
\label{fig:bigfgmaskingstudypspecs}
\end{figure*}

Using this estimator, we calculate the power spectrum after each stage of masking and plot the mean spectrum over all four deep ATLAS fields in Fig. \ref{fig:bigfgmaskingstudypspecs} (left panel). Instead of predicting the bandpower errors from the input covariances, we conservatively bootstrap the error bars by computing the standard deviation
 of each bandpower over the four fields. The power spectrum of all 850\,nm sources, masking only saturated regions, rises proportionally to $\ell$ (red dots), as expected for Poisson source counts when the power spectrum is plotted as $\Delta=\sqrt{\ell^2C_\ell/2\pi}$. Masking sources out to $5\sigma$ removes two orders of magnitude in power (green dots), and masking out to $12\sigma$ and above 70\,kJy/sr over the background removes a factor of a few more in power (blue dots). Lastly, excluding modes with $|\ell_x|<200$ or $|\ell_y|<200$ gives our final 850\,nm anisotropy spectrum (black dots). We note that more stringent masking does not significantly alter this final result. After the first masking stage, we use a flat power spectrum in $C(\ell)$ as our guess in the quadratic estimator formalism, and after the other three masking steps we use the 1.1\,$\mu$m power spectrum from \citet{zemcov14}.

In Fig. \ref{fig:bigfgmaskingstudypspecs} (right panel) we compare our final residual power spectrum measurement with other measurements\footnote{Note that $I_f/(\text{kJy/sr})\approx0.3\lambda I_\lambda/(\text{nW}/\text{m}^2/\text{sr})$ at $\lambda=1$\,$\mu$m.} in the literature. Our 850\,nm ATLAS spectrum (black dots) agrees very well with the ($1.1\pm0.25$)\,$\mu$m CIBER spectrum \citep{zemcov14} (blue dots), with much smaller error bars at the $\ell$ modes we can access ($\ell\lesssim4000$) due to ATLAS's larger field of view. 

\citet{zemcov14} argue that their spectrum is limited by foregrounds at all $\ell$: by diffuse Galactic light (DGL) (i.e., dust) at $\ell\lesssim1000$, by intrahalo light (IHL) (cyan dashed line) at $\ell\sim1000$, and by galaxy number counts below the flux limit at larger $\ell$. ATLAS's $2''$ resolution is only a factor of 3 finer than CIBER's $7''$ resolution, so perhaps it is not surprising that the level of galaxy number counts is not appreciably different. The ATLAS points at $\ell>1000$ are in fact slightly above CIBER's, though the measurements were conducted over slightly different bands. \citet{mw15} demonstrate a dramatic improvement in foreground reduction $\ell\gtrsim10^3$ in $10'$ Hubble mosaics with $0.1''$ resolution. Their power spectrum (red points) is a factor of $\sim10$ in amplitude below the previous IHL model, and their new fitting suggests that \citet{zemcov14} was in fact limited almost exclusively by DGL (red solid line). For comparison, we plot their \textit{new} IHL model (red dashed) and EOR model (grey area). These results show that finer angular resolution can help reduce the foregrounds in the $\ell\sim10^2-10^4$ modes that can be cross correlated with 21\,cm EOR intensity mapping experiments. 

\subsection{Modeling the 21\,cm--Ly$\alpha$ cross spectrum}
\label{sec:modelingthecrossspectrum}

\begin{figure*}[h]
\centering
\includegraphics[width=7in]{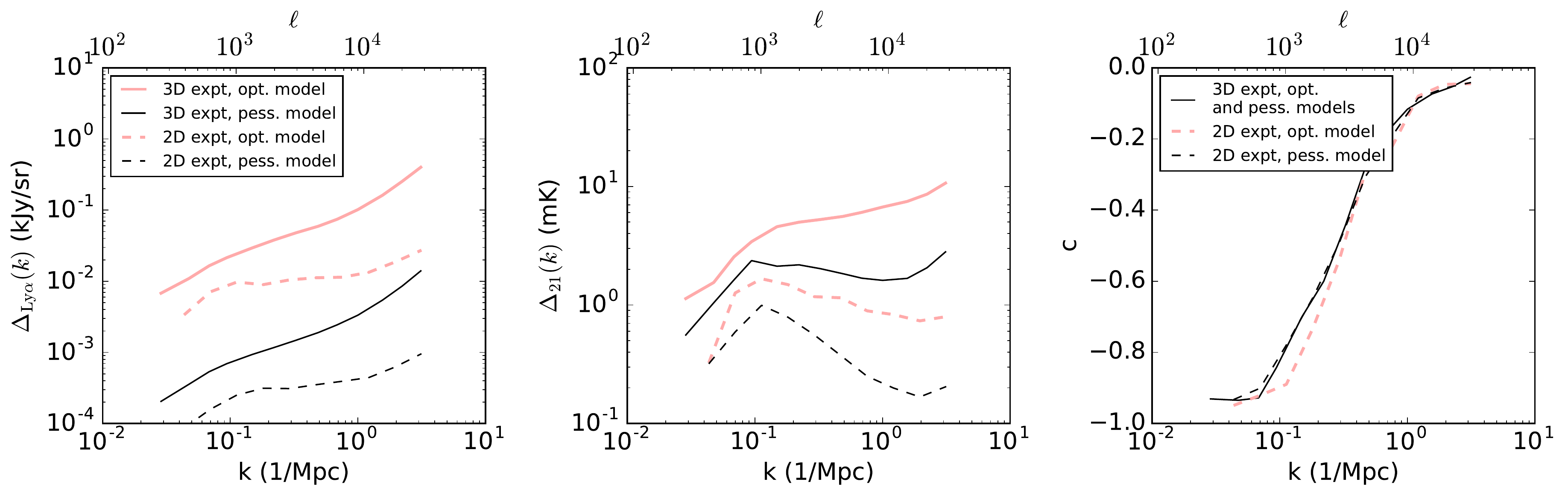}
\caption[Study of the relationship between power spectra recovered from 2D and 3D surveys, both for 21\,cm and 850\,nm cubes.]{As described in Sec. \ref{sec:modelingthecrossspectrum}, we identify optimistic (red solid lines) and pessimistic (black solid lines) Ly$\alpha$ (left panel) and 21\,cm (center panel) 3D power spectra from \citet{Gong2014} and \citet{PoberNextGen, 21cmfast}, respectively, and use the coherence function from \citet{Heneka2016} (right panel, black solid) for both scenarios. We then simulate approximate cubes assuming Gaussian statistics, average them over frequency, and plot the angular power spectra of Ly$\alpha$ (left panel, dashed lines) and 21\,cm emission (center panel, dashed lines), and their coherence functions (right panel, dashed lines). Power on short angular scales ($\ell\gtrsim10^4$) is suppressed by a factor of $\sim100$ (in power units), while power on longer angular scales ($\ell\lesssim10^2$) is suppressed by only a factor of $\sim2$ compared to the 3D power spectrum. This indicates that there remains significant observable signal in broad band 21\,cm--Ly$\alpha$ cross spectrum observations. }
\label{fig:spectra3Dto2D}
\end{figure*}

Before turning to 21\,cm--Ly$\alpha$ cross spectrum measurements with our 185\,MHz and 850\,nm images, we generate optimistic and pessimistic theoretical cross spectra for comparison. We simulate 21\,cm and Ly$\alpha$ cubes using 21\,cm power spectra from \citet{PoberNextGen}, Ly$\alpha$ power spectra from \citet{Gong2014}, and the coherence between the two fields from \citet{Heneka2016}. Combining simulations from all these sources allows us to better estimate the modeling uncertainty. Future work is needed to more self-consistently model these fields and their correlation over a range of possible reionization scenarios, and to infer astrophysical parameters from observed cross spectra.

\citet{Gong2014} model the Ly$\alpha$ cross spectrum and plot an uncertainty region over a range of likely values of escape fraction of ionizing photons, fraction of radiation emitted at Ly$\alpha$, star-forming rate, and IGM clumping factor (their Fig. 1). We take the upper and lower edges of their uncertainty region as our optimistic and pessimistic power spectra, respectively. 

Similarly, \citet{PoberNextGen} simulate 21\,cm power spectra over a range of reionization scenarios using \citet{21cmfast}, with various values of ionizing efficiency ($\zeta$) (including the escape fraction), the minimum halo virial temperature ($T_\text{vir}$), and the ionizing photon mean free path in the IGM ($R_\text{mfp}$). Whether the signal at our redshift of interest ($z=7$) is large or not depends mostly on whether reionization is already largely finished by that time or not. So we take as our pessimistic model the $z=8$ power spectrum of the ($\zeta =31.5$, $T_\text{vir}=1.5\times10^4$\,K, $R_\text{mfp}=30$\,Mpc) scenario, whose reionization midpoint is $z=9.5$. We take as our optimistic model a late reionization scenario with $T_\text{vir}=3\times10^5$\,K (other parameters unchanged), whose reionization midpoint is $z=5.5$. 

From these power spectra and coherence functions, we generate approximate 21\,cm and Ly$\alpha$ cubes assuming Gaussian statistics. This is an approximation, given that the 21\,cm field is expected to become less and less Gaussian as reionization proceeds \citep{skew}, and more work is needed to understand this effect. The simulated cubes have (1\,Mpc)$^3$ resolution over a (218\,Mpc)$^3$ volume at $z=7$, corresponding to $\Delta z=0.6$ and $0.4'$ angular resolution. Our 21\,cm and Ly$\alpha$ cubes have units of mK and kJy/sr, respectively, and we average them in the line of sight direction to produce broad band images.

We plot the original 3D spherically averaged power spectra and coherence function as solid lines in Fig. \ref{fig:spectra3Dto2D}, and the computed 2D power spectra from the line-of-sight averaged cubes, as well as their coherence as dashed lines. The 3D and 2D power spectra are plotted as $\Delta(k)$ and $\Delta(\ell)$, respectively, as defined in Sec. \ref{sec:pspecdefs}. As expected, line of sight averaging tends to remove power on short spatial scales (compared to the line of sight depth of the cube), but it acts similarly on both cubes, so the coherence function is preserved. Note that the cross spectrum $\Delta_{12}$ is defined in terms of the coherence as $\Delta^2_{12}=c[\Delta^2_1\Delta_2^2]^{1/2}$ for both the 3D and 2D cases, and we use the same coherence function from \citet{Heneka2016} in the optimistic and pessimistic scenarios. These results show that 2D experiments can detect much of the 3D fluctuation power at $k<0.2$\,Mpc$^{-1}$, and motivate more complete and self consistent simulations of the 21\,cm and Ly$\alpha$ fields throughout the EOR over a range of optimistic and pessimistic scenarios.

\subsection{Limits on the 21\,cm--Ly$\alpha$ cross spectrum and an experimental design study}
\label{sec:limitsandexptdesignstudy}

Having characterized the residual sky power spectrum at 185\,MHz and 850\,nm after applying the best foreground masking and subtraction permitted by our datasets, we now search for a correlation between these radio and foreground residuals. To account for the non-uniform $uv$ sampling of the MWA image as well as for the image space masking of the ATLAS image, we again use the optimal quadratic estimator. In Appendix \ref{sec:optimalestimatorforcrossspectrum} we show that with the approximation that the correlation between the two images is small, the proper extension of the optimal quadratic estimator from the auto spectrum case to the cross spectrum case is given by
\begin{equation}
q_\alpha \approx (\xb_{21}-\langle\xb_{21}\rangle)^t \Cb_{21}^{-1} \Cb_{,\alpha}\Cb_\IR^{-1}(\xb_\IR-\langle\xb_\IR\rangle)
\end{equation}
\begin{equation}
\Fb_{\alpha\beta}\approx\text{tr}\left(\Cb_{21}^{-1} \Cb_{,\alpha} \Cb_\IR^{-1}  \Cb_{,\beta}  \right)	
\end{equation}
where $\Cb_{,\alpha}$ is the same matrix used in Sec. \ref{sec:resirfg}, and $\mathbf{x}_{21}$ and $\mathbf{x}_{\IR}$ are column vectors of 21\,cm and IR pixel values. Observe that in this approximation, we model the radio covariance separately from the IR covariance. This separation is also ideal for our current study where we seek to \textit{characterize} any cross spectrum between the two, rather than \textit{model} a cross spectrum and \textit{downweight} by it.

As before we use normalization $\Mb=\Fb^{-1/2}$ in $\pb=\Mb\qb$, and perform a final normalization using the peaks of the window functions $\Wb=\Mb\Fb$. We use the same IR covariance matrix given in Eqn. \ref{eqn:covFTwithmask}, and model the radio covariance matrix as solely due to sample variance. We construct this covariance using the first term on the right side of Eqn. \ref{eqn:covFTwithmask}, taking the power spectrum of the residual MWA broadband images as our guess. We emphasize that in both cases, random noise is subdominant to the foreground residuals, so we do not include photon shot noise or thermal noise in these covariances. 

\begin{figure}[h]
\centering
\includegraphics[width=3.5in]{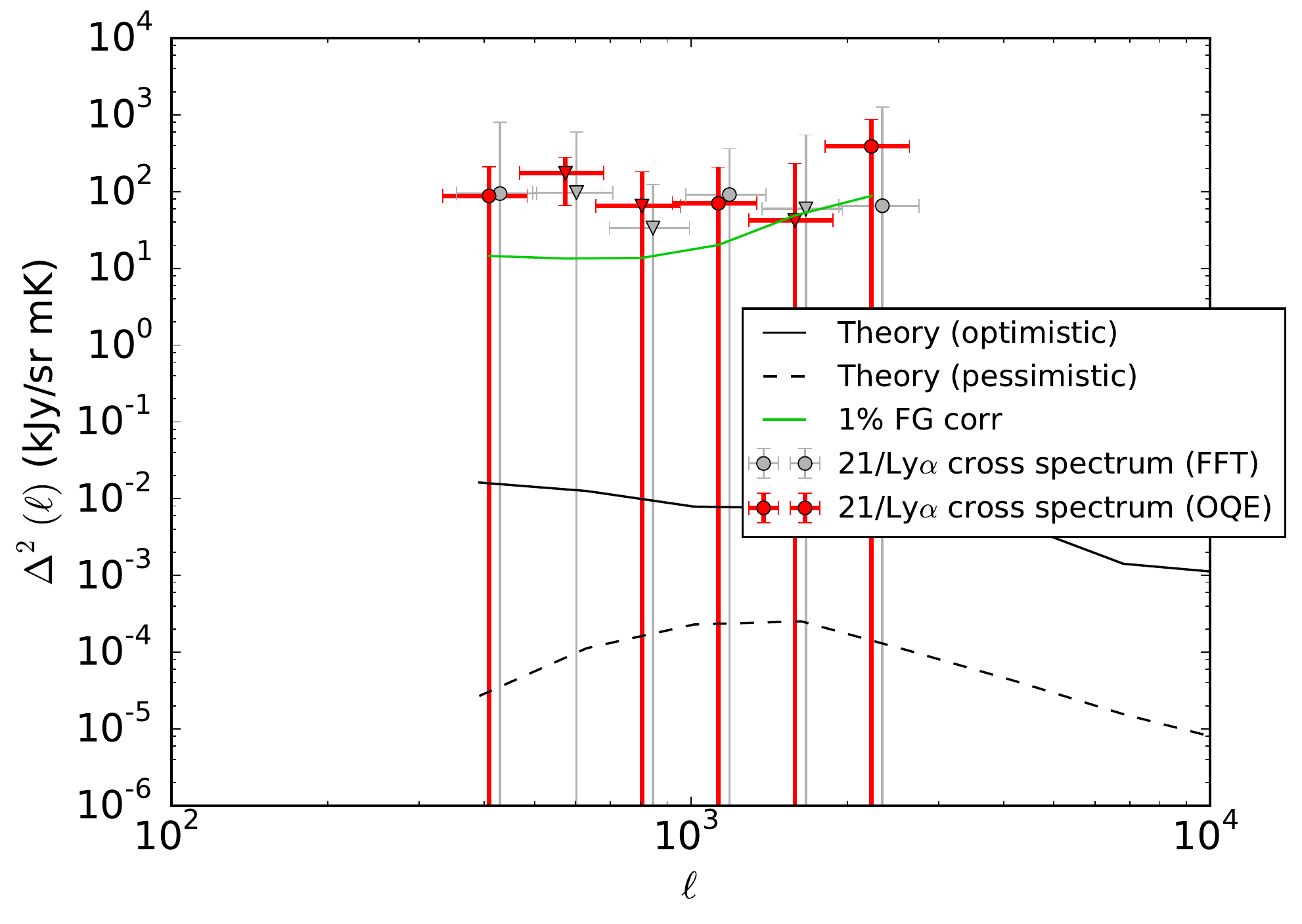}
\caption[Measured cross spectrum between 185\,MHz and 850\,nm images after our best foreground subtraction and masking.]{Measured cross spectrum between 185\,MHz and 850\,nm images after our best foreground subtraction and masking. Our optimal quadratic estimator results are shown as red markers, while our FFT-based cross spectrum results are shown as grey markers. Most are consistent with zero, and roughly half are positive (circles) and negative (triangles), as expected if there is no measured correlation. Error bars show $2\sigma$ uncertainties, and we take the tops of these error bars as upper limits on the cross spectrum of residual foregrounds of 21\,cm--Ly$\alpha$ emission from the EOR at $z\sim7$. We achieve an upper limit of $\Delta^2<181$\,$(\text{kJy/sr}\cdot \text{mK})$ (95\%) at $\ell\sim800$. We also plot the radio--infrared foreground cross spectrum that a 1\% flux correlation would produce (green line), and find that it is below the sensitivity of our MWA--ATLAS experiment, which is  limited by the sample variance noise of uncorrelated foregrounds.}
\label{fig:resxspec}
\end{figure}

We apply this formalism to each of the four $4^\circ$ deep ATLAS fields shown in Fig. \ref{fig:surveyoverview}, pairing each IR image with the overlapping region of the MWA image cropped from the naturally weighted image. We crop the image space synthesized beam over the same field of view, then use it to apply uniform weighting to the cropped MWA image using Eqn. \ref{eqn:uniformweighting}. In Fig. \ref{fig:resxspec} we plot the resulting cross spectrum (red markers) averaged over all four fields with $2\sigma$ error bars. Most are consistent with zero, and roughly half the estimated bandpowers are negative (triangles) and positive (circles), as expected if there is no measured correlation. Our bins are evenly spaced in $\log \ell$, and it is beneficial to overlap them slightly since our normalization matrix $M\sim F^{-1/2}$ ensures bin errors are always uncorrelated. We take the tops of the error bars as 95\% upper limits on the cross spectrum of residual foregrounds of  21\,cm and Ly$\alpha$ emission from the EOR, achieving a tightest upper limit of $\Delta^2<181$\,$(\text{kJy/sr}\cdot \text{mK})$ (95\%) at $\ell\sim800$. For comparison, we plot optimistic (black solid) and pessimistic (black dashed) model cross spectra as discussed in the previous section.
 
We check these quadratic estimator results (red points) against FFT-based power spectrum results on the same images (grey points). In fact the two spectra are quite similar. In most bins the quadratic estimator limits are lower, but are not significantly closer to the fiducial 1\% radio/IR foreground correlation. To understand why the FFT results are only slightly worse, consider that after our IR foreground masking, only $\sim$5\% of coarse pixels end up masked. With different masking parameters more or fewer coarse pixels would get masked, in some cases necessitating the quadratic estimator to deal with large masked regions. But with the best masking parameters we settled on, the improvement is modest.

Also in Fig. \ref{fig:resxspec} we plot the radio--infrared foreground cross spectrum that a 1\% flux correlation would produce (green line), as motivated by our results in Sections \ref{sec:fluxcorsims} and \ref{sec:catcorrelations}. Such a slight geometric flux correlation is thus still predicted to be slightly below the sensitivity of our MWA--ATLAS experiment, but is it significant for other classes of experiments? To study this we predict for a range of current and future radio--IR surveys both the foreground sample variance noise and the foreground cross spectrum for a slight geometric correlation.

\begin{table}
\caption{Current and future radio and IR surveys\label{tab:expts}}
\begin{tabular}{|l | l | l|}
\hline
\textbf{Radio Survey} & \textbf{Resolution} & \textbf{$S_\text{max}$} \\
\hline
MWA & $6'$ & 150 mJy \\
MWA (GMRT catalog) & $6'$ & 25 mJy \\
HERA (GMRT catalog) & $12'$ & 25 mJy \\
GMRT & $40''$ & 25 mJy \\
LOFAR & $20''$ & 3 mJy \\
SKA & $10''$ & 0.1 mJy \\
\hline\hline
\textbf{IR Survey} & \textbf{Field of View} & \textbf{$f_\ir$}\footnote{Fraction of IR foreground power remaining relative to our ATLAS analysis.} \\
\hline
ATLAS & $8^\circ$ & 1.0 \\
WideATLAS & $40^\circ$ & 1.0 \\
WFIRST mosaic & $4^\circ$ & 0.01 \\
DES mosaic & $40^\circ$ & 0.01 \\
\hline
\end{tabular}
\end{table}

We describe the radio and IR surveys we study here, and summarize them side by side in Table \ref{tab:expts}. \\

\noindent\textbf{Radio surveys}
\begin{itemize}
\item \textit{MWA}. We use the $6'$ resolution permitted by our MWA analysis, corresponding to a maximum baseline length of 700\,m, and a survey depth at 50\% completeness of 150\,mJy (see Sec. \ref{sec:catalogs}).
\item \textit{HERA}. Primarily designed for 3D power spectrum measurements which are severely thermal noise limited, HERA is a compact redundant array, and even the outriggers only improve the resolution to $12'$ \citep{deboer16}. We assume the GMRT catalog is used for source subtraction (see below) .
\item \textit{GMRT}. We assume a maximum baseline length of 10\,km, within which the $uv$ plane is filled relatively uniformly, this corresponds to a $40''$ resolution. \citet{intema17} demonstrate a catalog depth of 25\,mJy at 50\% completeness.
\item \textit{LOFAR}. Similarly, we use a maximum baseline length of 20\,km, to maintain a relatively filled $uv$ plane, corresponding to a $20''$ resolution. \citet{lofareorpaper} report a catalog depth of 3\,mJy. 
\item \textit{SKA}. \citet{prandoni15} report a nominal SKA-Low confusion-limited survey depth of 0.1\,mJy at $10''$ resolution.
\end{itemize}

\noindent\textbf{IR surveys}
\begin{itemize}
\item \textit{ATLAS}. Our four ATLAS stacking field give a total field of view of $8^\circ$, though we are planning a future wider survey with a $40^\circ$ field of view. 
\item \textit{WFIRST mosaic}. \citet{mw15} have demonstrated high fidelity mosaicing using the technique of \citet{fixen00}, producing $10'$ wide mosaics from the $\sim2'$ Hubble field of view. Along these lines, we assume that a $4^\circ$ WFIRST field can be used through some combination of mosaicing (making larger IR images) and tiling (averaging the cross spectrum over many small fields). The instantaneous field of view of WFIRST is roughly 100 times that of Hubble with comparable resolution. Motivated by the results of \citet{mw15}, we assume a 100$\times$ improvement in IR foreground removal over our ATLAS analysis.
\item \textit{DES mosaic}. Along these lines, we assume a similar analysis can be conducted using the Dark Energy Survey \citep{des16} over a field of view which is 10$\times$ larger on a side, due to the instrument's correspondingly larger field of view. We assume the same improvement in foreground removal as in the Hubble mosaic discussed previously.
\end{itemize}

Neglecting the small geometric foreground correlation, the foreground sample variance noise due to uncorrelated foregrounds is
\begin{eqnarray}
\Delta^2_\text{noise}(\ell) &=& \sqrt{\frac{\Delta^2_\IR(\ell) \Delta^2_\rad(\ell)}{2N_\ell}} \nonumber \\
&=&\sqrt{\frac{\Delta^2_{\text{ATLAS}}(\ell) \Delta^2_{\text{MWA}}(\ell)f_\rad f_\ir}{2N_\ell}} \nonumber \\
&&\times \left( \frac{\theta_{\fov,\text{ATLAS}}}{\theta_\fov}\right)\left(\frac{d\theta}{d\theta_{\text{MWA}}}\right)
\end{eqnarray}
where $f_\ir$ and $f_\rad$ are the fractions of IR and radio foreground power remaining relative to those in our ATLAS and MWA analyses. Using the empirical formula for radio source counts given in \citet{dimatteo02}, $f_\rad$ scales as $f_\rad\propto S_\text{max}^{1.25}$, where $S_\text{max}$ is the flux limit of the subtraction. Scaling this relative to our MWA analysis gives $f_\rad=(S_\text{max}/S_{\text{max},\text{MWA}})^{1.25}$.

Similarly, the foreground cross spectrum for a correlation $c$ is given by
\begin{eqnarray}
\Delta^2_\text{FG}(\ell)& =&c\sqrt{\Delta^2_\IR(\ell) \Delta^2_\rad(\ell)} \nonumber \\
&=&c\sqrt{\Delta^2_{\text{ATLAS}}(\ell) \Delta^2_{\text{MWA}}(\ell)f_\rad f_\ir }
\end{eqnarray}

\begin{figure}[h]
\centering
\includegraphics[width=3.5in]{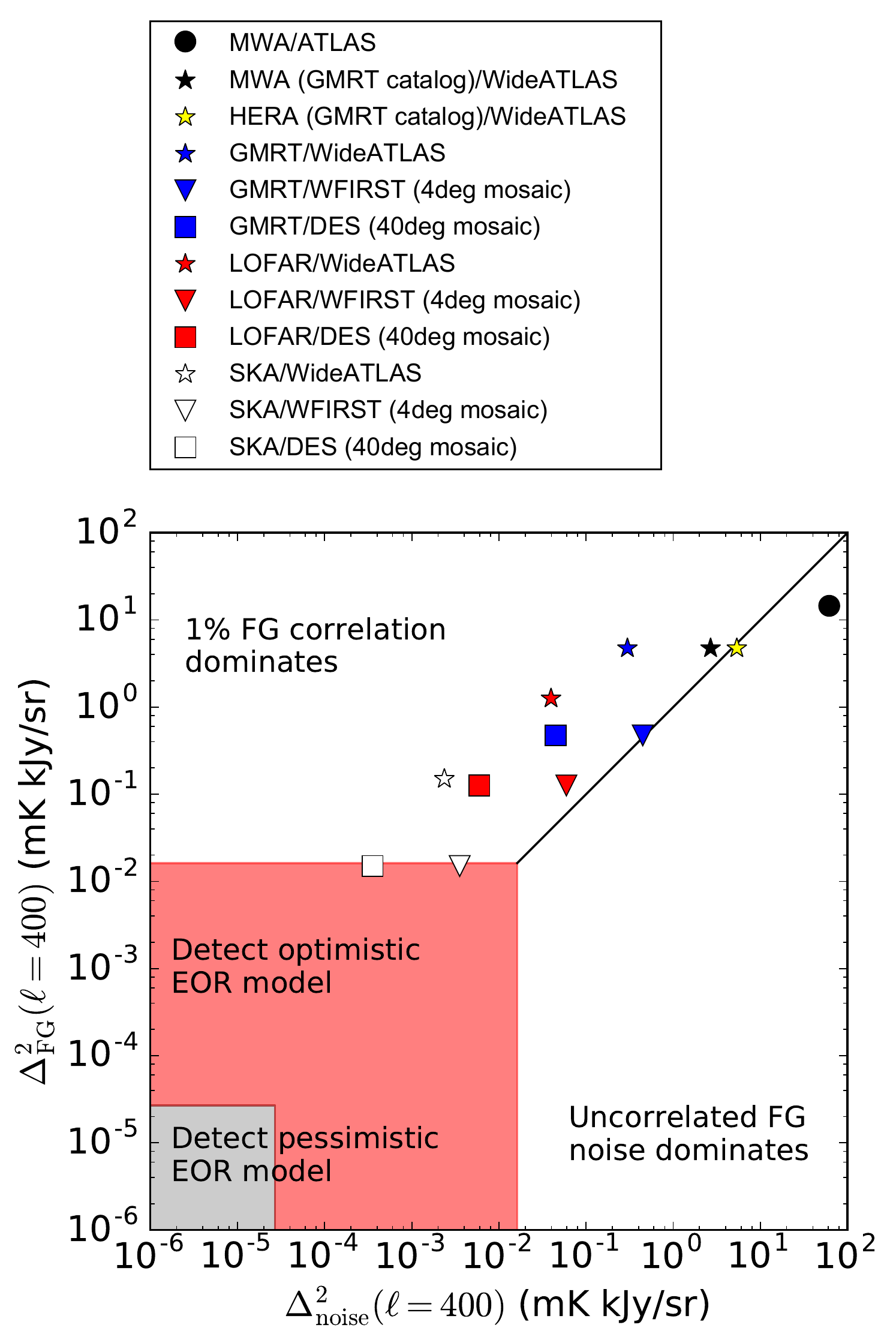}
\caption[Experimental design study for 21\,cm--Ly$\alpha$ cross spectrum measurements, weighting foreground cross spectra due to flux correlatians against foreground sample variance.]{Foreground cross spectrum due to percent-level geometric correlations (y-axis) due to sample variance noise due to the uncorrelated component of foregrounds (x-axis) for a number of possible radio--IR intensity mapping survey pairs. This work (black circle) is the only one limited by sample variance noise, even a nominal improvement using the deeper GMRT catalog becomes foreground correlation-limited. Deeper radio (e.g., LOFAR and SKA) and IR (e.g., Hubble and DES mosaics) are needed to give the foreground removal required to suppress the geometric foreground correlation below the optimistic hexpected EOR signal.}
\label{fig:noisecorrstudy}
\end{figure}

In Fig. \ref{fig:noisecorrstudy} we plot the foreground sample variance noise versus foreground cross spectrum for a range of current and future radio and infrared survey pairs. We calculate these for the $\ell=400$ bin and assume a 1\% geometric foreground correlation. The diagonal black line separates the regions within which each of these effects dominates. We observe that our MWA/ATLAS analysis (black circle) is the only experiment limited by sample variance noise of uncorrelated foregrounds. An improved analysis using the GMRT catalog for radio source subtraction from MWA images and widening the ATLAS survey to a wide $40^\circ$ (black star) falls barely in the foreground correlation limited regime. A similar analysis using HERA instead of the MWA (yellow star) produces similar results, as HERA is not optimized for the wide field, high resolution radio surveys that broad band correlation experiments require. 

LOFAR reaches within a factor of a few in power of the optimistic EOR cross spectrum model when correlated against the Hubble (red triangle) or DES (red square) mosaics, and the SKA coupled with the same IR surveys achieves a near detection of the optimistic model (white square and white triangle). None of the experiments reaches the pessimistic model, but beginning to probe optimistic models already starts to exclude late reionization scenarios, and performing the same correlation experiment at higher redshifts can likely give more leverage on earlier ones. 

Interestingly, we note that virtually all the intermediate experiments between this work and the ultimate detection will be limited by the geometric foreground correlations. Motivated by our results in Sec. \ref{sec:catcorrelations}, we have assumed here that the coherence remains at the percent-level irrespective of mask depth, though more work is needed to model this effect more completely and self-consistently. In any case, this geometric foreground correlation appears to be strictly positive, while the predicted correlation of 21\,cm--Ly$\alpha$ from the EOR is negative, giving us a test of whether foregrounds remain in the data.

\section{Discussion}

Radio and infrared surveys are on the cusp of direct observations of the two sides of the EOR. Deep infrared surveys are beginning to probe the bright end of the ionizing galaxy population itself, while low frequency radio surveys are constraining the 21\,cm brightness of the neutral IGM. New radio and infrared instruments are all but assured to yield the first direct constraints on the astrophysics of the EOR, and combining these measurements will be crucial to confirm these first detections and extend their science reach. In particular, measurements of the predicted anticorrelation between redshifted radio and Ly$\alpha$ emission can shed light on properties and statistics of the ionizing sources which 21\,cm measurements are not directly sensitive to, while simultaneously yielding redshift information on the near infrared background for the first time. 

3D correlation measurements are the ultimate goal, but making infrared maps with sufficient spectral resolution over large enough fields to match the comoving volumes probed by low frequency 21\,cm experiments is challenging and expensive. In contrast, 2D (i.e., broad band) infrared maps can be measured by many existing ground- and space-based observatories, and are a natural first step. 

We have shown that foregrounds pose two significant challenges for 2D intensity mapping correlation experiments: (1) simple geometric effects result in percent-level correlations between radio and infrared fluxes, even if their luminosities are uncorrelated; and (2) the largely uncorrelated radio and infrared foreground residuals contribute a sample variance noise to the cross spectrum. The first challenge demands better foreground masking and subtraction, while the second requires measurements over large fields of view with many independent samples to average down uncorrelated radio and infrared power.

In the first part of this paper we searched for correlations between radio and infrared catalog fluxes (i.e., point sources). We began with the 185\,MHz MWA catalog and the 4.5\,$\mu$m WISE catalog; any foreground correlation between these bands would limit 21\.cm--H$\alpha$ correlation analyses at $z\sim7$. We detect a few correlation at $>6\sigma$ significance after masking 4.5\,$\mu$m sources down to 18.25 mag, which corresponds to the flux below which extragalactic sources dominate the catalog. We reproduce this observed 185\,MHz--4.5\,$\mu$m correlation in simulations, confirming that it is sourced by starforming galaxies rather than AGN. The narrow radio and infrared luminosity functions of the former make distance a much stronger effect than for AGN in deriving fluxes from luminosities. 

We then performed 850\,nm observations of the MWA field using the ATLAS telescope, and searched for correlations between 185\,MHz and 850\,nm catalog fluxes in MWA and ATLAS catalogs. Any correlation between these bands would limit 21\,cm--Ly$\alpha$ analyses. We observed a marginal correlation at the $\sim3\sigma$ level which grew slightly stronger towards our deepest 850\,nm mask depth of 0.1\,mJy. This is consistent with our finding from Sloan Digital Sky Survey data that at 850\,nm, stars still dominate source counts above this flux. Deeper near infrared studies are needed to study how these correlations depend on mask depth to quantify what level of foreground removal is required to mitigate them. For now, we assume that radio--infrared flux correlations are at the percent-level for the purpose of weighing foreground flux correlations versus foreground sample variance noise in our experimental design study.

In the second part of this paper we analyzed the residual 185\,MHz and 850\,nm foregrounds in MWA and ATLAS images after subtracting radio sources and masking infrared sources. We computed power spectra of the band averaged MWA cubes and found they agree with the $k_\parallel=0$ bin of the 3D power spectra of \citet{beardsley16} after conversion from 3D to 2D units. Despite the fact that the MWA image is not thermal noise-limited, we find that foreground subtraction improves significantly (by a factor of 2 in power) when the same number of observations are spread over two weeks instead of clustered within a single night, consistent with ionosphere-induced calibration errors. If the ionosphere is indeed the cause, then the implication is that is that observing time should be rotated between multiple fields on a nightly cadence.

We then used the optimal quadratic estimator of+ \citet{Maxpowerspeclossless} to compute the power spectrum of our 850\,nm ATLAS images after masking sources at $2''$ resolution. These results agree well over $300<\ell<4000$ with the $1.1$\,$\mu$m power spectrum measured by \citep{zemcov14} using CIBER, a sounding rocket with $7''$ resolution and $\sim1^\circ$ field of view. Our larger field of view permits significantly improved precision at $\ell<1000$, and while our slightly increased resolution \textit{should} permit better masking, the lack of foreground reduction indicates that the power spectrum has hit a floor at these angular scales. \citet{mw15} show that mild improvement is possible using the Hubble's significantly improved $0.1''$ resolution, but argue that further improvement at these angular scales is stymied by Galactic dust. They demonstrate that significant improvement is possible at larger $\ell$, though these scales are largely inaccessible to 21\,cm observatories focusing on EOR power spectrum measurements (e.g., PAPER, the MWA, and HERA). Taking advantage of these infrared foreground reductions will require the higher resolution images of  LOFAR and SKA-LOW.

Turning to cross spectrum measurements, we simulate the loss of signal in to band averaging from 3D to 2D maps. We find that 2D experiments can recover much of the signal at $k<0.2$\,Mpc$^{-1}$, but that power on shorter length scales  suppressed by a factor of $10-10^2$ in power due to averaging out of of $k_\parallel>0$ modes. 

Finally, we used our foreground subtracted and masked MWA and ATLAS images to set the first limits on the cross spectrum of residual foregrounds of 21\,cm and Ly$\alpha$ emission  from $z\sim7$. We adapted the optimal quadratic estimator to the cross spectrum case to properly account for both non-uniform radio $uv$ sampling and infrared image space masking. Our results are consistent with zero correlation, and our strictest upper limit on the residual foreground cross spectrum is $\Delta^2<181$\,$(\text{kJy/sr}\cdot \text{mK})$ (95\%) at $\ell\sim800$.

A percent-level correlation between radio and infrared foreground fluxes, of the sort suggested by our catalog correlation analyses, remains below this upper limit, but will be crucial for future experiments. We weigh the impact of a percent-level radio--infrared foreground correlation against the sample variance noise due to their uncorrelated part for several possible future experiments. The simplest improvement we consider is increasing the ATLAS survey area by a factor of $\sim10$ and subtracting radio fluxes below the MWA's confusion limit using the GMRT catalog. We predict that the sensitivity boost of these improvements should reveal the cross spectrum floor due to the percent-level geometric flux correlations. Pushing down this floor requires foreground reduction using to higher resolution radio surveys such as LOFAR and SKA-LOW and infrared surveys such as Hubble and the Dark Energy Survey, though future work is needed to better understand these geometric flux correlations at very deep flux cuts.

We conclude that detection of the predicted EOR anticorrelation between redshifted 21\,cm and Ly$\alpha$ emission in 2D (i.e., broad band) images is challenging, but within reach of future surveys.

In the near term, by probing correlation properties of radio/infrared foregrounds, these 2D experiments will be valuable complements to future 3D correlation analyses using infrared cubes from the proposed SPHEREx and Cosmic Dawn Intensity Mapper telescopes. These 3D surveys exchange foreground challenges for sensitivity and cost challenges, but will eventually probe the short spatial scales inaccessible to 2D surveys. The decade of direction observation of the EOR is upon us, and correlation experiments will help us move from the era of detection to the era of astrophysics.

\begin{acknowledgments}
We acknowledge helpful discussions on optimal quadratic estimators with Adrian Liu, Josh Dillon, and Aaron Ewall-Wice, and discussions on MWA image products from the FHD pipeline with Adam Beardsley, Bryna Hazelton, Danny Jacobs, and Nichole Barry. 
\end{acknowledgments}

\appendix

\section{Power spectrum of photon shot noise}
\label{sec:Pshot}

In Sec. \ref{sec:resirfg} we measure the maximum airglow to be $I_\text{air}=5\times10^3$ kJy/sr, and in this appendix we calculate the power spectrum of this photon shot noise. We must observe that the mean number of photons collected by a pixel during each observation is $\langle N_\text{ph}\rangle=I_\text{air}At_\text{int} \Delta f d\theta^2/hf$, where $A=(0.5\,\text{m})^2$ is the collecting area of ATLAS, $t_\text{int}=30\,$sec, $\Delta f$ and $f$ are the frequency bandwidth and center frequency of I band, and $d\theta$ is the pixel size. The passband has $\Delta\lambda=150\,$nm and $\lambda=800\,$nm. 

The shot noise contribution to the power spectrum is given by
\begin{equation}
C_{\IR, \shot}(\vec{\ell}) = \left\langle\left|\sum_{m,n}I_\shot(m,n)e^{-2\pi i(ma+nb)/N}\right|^2\right\rangle \frac{d\theta^2}{N^2}
\end{equation}
where $I_\shot(m,n)\equiv I(m,n)-\langle I(m,n)\rangle$ denotes the photon shot noise contribution to pixel (m,n), and $N$ is the number of pixels on each side of the square image. Then using the fact that the shot noise is uncorrelated between different pixels, we find
\begin{equation}
C_{\IR, \shot}(\vec{\ell}) = \sum_{m,n}\langle I^2_\shot(m,n)\rangle \frac{d\theta^2}{N^2}
\end{equation}
Note that $I(m,n)=N_\text{ph}(m,n)hf/\Delta f A t_\text{int}d\theta^2$ and $\langle N_\text{ph}^2\rangle = \langle N_\text{ph}\rangle$, so we have
\begin{equation}
C_{\IR, \shot}(\vec{\ell}) = \langle N_\text{ph}\rangle \left(\frac{hf}{\Delta f A t_\text{int}d\theta^2}\right)^2 d\theta^2
\end{equation}

\begin{equation}
C_{\IR, \shot}(\vec{\ell}) =\frac{I_\text{air}h\lambda}{\Delta \lambda A t_\text{int}}
\end{equation}
which gives $\Delta(\ell=10^3)=6\times10^{-2}$\,kJy/sr for our deep fields with $t_\text{int}=2.5$\,min. 

\section{Relation between the power spectrum of image cubes and broadband images}
\label{sec:pspecrelation}

We focus in this paper on the spherical power spectrum of broadband images, $C_\ell$,  instead of that of image cubes, $P(\vec{k})$, as 21\,cm observations have focused on. Here we work out the approximate relation between the two over small fields of view (i.e., for large $\ell$) to facilitate comparison with past 21\,cm power spectrum results. In particular, we calculate the scaling factor $B$ relating the purely transverse modes of the power spectrum $P(k_\perp,k_\parallel=0)$ of a image cube $I(\theta_x,\theta_y,f)$ to the spherical power spectrum of a broad band image $C_\ell$ as
\begin{equation}
P(k_\perp,k_\parallel=0) = B C_{\ell(k_\perp)}
\end{equation}

Using the Fourier transform convention discussed in Sec. \ref{sec:pspecconventions}, the left side of the equation is given by
\begin{equation}
P(k_\perp,k_\parallel=0) = \frac{1}{N_\perp^2 N_\parallel dV}\langle|\tilde{I}(k_x,k_y,k_\parallel=0)|^2\rangle
\end{equation}
where $N_\perp\equiv N_x=N_y$ is the number of pixels in each of the two transverse dimensions of the image cube, and $N_\parallel$ is the number of pixels in the line of sight (ie, frequency) dimension. The comoving pixel volume is $dV = (D_c d\theta)^2 (\Delta D_c/N_\parallel)$, where $D_c$ is the line of sight comoving distance from the present day to the center of the cube, and $\Delta D_c$ is the comoving line of sight thickness of the cube. Lastly, recall that $k_\perp$ is related to $k_x$ and $k_y$ as $k_\perp=\sqrt{k_x^2+k_y^2}$.

Now substituting the definition of the Fourier transform, we find

\begin{equation}
P(k_\perp,k_\parallel=0) =\frac{1}{N_\perp^2 N_\parallel dV}\left\langle\left|dV\sum_{\theta_x,\theta_y,f}I(\theta_x,\theta_y,f)e^{iD_c(k_x\theta_x+k_y\theta_y)}\right|^2\right\rangle
\end{equation}

Simplifying and writing this in terms of the broadband image $I_{\Delta f}(\theta_x,\theta_y)\equiv\frac{1}{N_\parallel}\sum_f  I(\theta_x,\theta_y,f)$, we find

\begin{equation}
P(k_\perp,k_\parallel=0) =(D_c^2 \Delta D_c)
\frac{d\theta^2}{N_\perp^2}\left\langle\left|\sum_{\theta_x,\theta_y}I_{\Delta f}(\theta_x,\theta_y)e^{iD_c(k_x\theta_x+k_y\theta_y)}\right|^2\right\rangle
\end{equation}

Now denote $k_x=a\cdot dk$, $k_y=b\cdot dk$, $\theta_x=m\cdot d\theta$, and $\theta_y=n\cdot d\theta$, where $dk = 1/N_\perp D_c d\theta$. 

\begin{equation}
P(k_\perp(\ell(a,b)),k_\parallel=0) =(D_c^2 \Delta D_c)
\frac{d\theta^2}{N_\perp^2}\left\langle\left|\sum_{m,n}I_{\Delta f}(m,n)e^{2\pi i(am + bn)/N_\perp}\right|^2\right\rangle
\end{equation}

Comparing with Equations \ref{eqn:Cldef}, \ref{eqn:elldef}, and \ref{eqn:elldef2}, we see that $B\equiv P(k_\perp,k_\parallel=0)/ C_{\ell(k_\perp)}=D_c^2 \Delta D_c$ and $\ell=D_c k_\perp$.

\section{Extending the optimal quadratic power spectrum estimator to the cross spectrum case}
\label{sec:optimalestimatorforcrossspectrum}

The optimal quadratic estimator formalism presented in Sec. \ref{sec:resirfg} was constructed to estimate the power spectrum of an image with arbitrary pixel sampling and noise properties. In this section we extend this formalism to achieve the same advantages in cross spectrum measurements. 

Consider measurements at two bands over the same set of pixels on the sky, $\xb_1$ and $\xb_2$, each column vectors with $N$ elements. Let us combine these together into a single column vector containing all measurements as $\xb=\left(\begin{matrix}\xb_1 \\ \xb_2  \end{matrix}\right)$, whose covariance is given by 

\begin{equation}
\Cb\equiv  \langle\xb\xb^\dagger\rangle-\langle\xb\rangle\langle\xb^\dagger\rangle=\left(\begin{matrix}\Cb_1 & \Cb_{12} \\ \Cb_{12}^\dagger & \Cb_2   \end{matrix}\right)
\end{equation}
and 
\begin{equation}
\frac{d\Cb}{dp_{12}^\alpha}=\left(\begin{matrix}\mathbf{0} & \Cb_{,\alpha}\\ \Cb_{,\alpha}^\dagger & \mathbf{0}   \end{matrix}\right)
\end{equation}
$\Cb_1$ and $\Cb_2$ depend only on the auto power spectra of the different fields; only $\Cb_{12}$ depends on the cross spectrum. Said another way, $C_{,\alpha}$ is the same matrix used in Sec. \ref{sec:resirfg}, but used here in the off-diagonal parts of $d\Cb/dp_{12}^\alpha$ so as to capture the cross products between the two fields\footnote{One might object to our form of $d\Cb/dp_{12}^\alpha$, arguing that artificially limiting ourselves to cross products between the two fields is tantamount to throwing a significant amount of the information contained in the data sets, and thus our estimator cannot be optimal. There are certainly situations in which this would be the case. If we had some theory of how each field was related to the matter density field of the universe, then both the auto-products and cross-products contain similar information. However we take an empirical approach where we assume we know nothing about either field and want only to know their correlation properties. Only the cross products contain that information.}. And as before, the unnormalized estimator $q_\alpha$ of the power in band $\alpha$ is given by
\begin{equation}
q_\alpha = \frac{1}{2}(\xb-\langle\xb\rangle)^t \Cb^{-1} \frac{d\Cb}{dp_{12}^\alpha}\Cb^{-1}(\xb-\langle\xb\rangle)
\end{equation}
and the elements of the Fisher matrix are
\begin{equation}
\Fb_{\alpha\beta}=\frac{1}{2}\text{tr}\left(\Cb^{-1} \frac{d\Cb}{dp_{12}^\alpha} \Cb^{-1}  \frac{d\Cb}{dp_{12}^\beta}  \right)	
\end{equation}

In the case where the correlation between the two fields is expected to be weak, and we are primarily interested in setting an upper limit, we can get a significant speedup by approximating $\Cb_{12}\approx0$ in our guess covariance. The expressions for $q_\alpha$ and $F_{\alpha\beta}$ then simplify to:
\begin{equation}
q_\alpha \approx (\xb_1-\langle\xb_1\rangle)^t \Cb_1^{-1} \Cb_{,\alpha}\Cb_2^{-1}(\xb_2-\langle\xb_2\rangle)
\end{equation}
\begin{equation}
\Fb_{\alpha\beta}\approx\text{tr}\left(\Cb_1^{-1} \Cb_{,\alpha} \Cb_2^{-1}  \Cb_{,\beta}  \right)	
\end{equation}





\end{document}